\documentclass[%
 aip,
 jcp,
 amsmath,amssymb,
 reprint,%
 floatfix,
]{revtex4-1}
\usepackage[utf8]{inputenc}
\usepackage{amsmath}
\usepackage{graphicx}   
\usepackage{bm}
\usepackage{hyperref}
\usepackage{color}
\usepackage{array}
\usepackage{siunitx}
\usepackage{mathtools}
\usepackage{cancel}
\usepackage[version=3]{mhchem}
\usepackage[normalem]{ulem}
\graphicspath{{./}{./Figures/}}

\let\jcp\undefined

\newcommand{\cpc}{Comput.\ Phys.\ Commun.\ }

\newcommand{\jcp}{J.\ Chem.\ Phys.\ }

\newcommand{\jctc}{J.\ Chem.\ Theory Comput.\ }
\newcommand{\jpca}{J.\ Phys.\ Chem.\ A }
\newcommand{\jpcb}{J.\ Phys.\ Chem.\ B }

\newcommand{\pccp}{Phys.\ Chem.\ Chem.\ Phys.\ }

\newcommand{\pnas}{Proc.\ Natl.\ Acad.\ Sci.\ }
\newcommand{\wn}{\ensuremath{\mathrm{cm}^{-1}}}

\newcommand{\fig}[1]{Figure~\ref{#1}}
\newcommand{\sfig}[1]{Fig.~\ref{#1}} 

\newcommand{\lsec}[1]{Section~\ref{#1}} 
\newcommand{\ssec}[1]{Sec.~\ref{#1}} 
\newcommand{\ssecc}[2]{Sec.~\ref{#1} and~\ref{#2}} 
\newcommand{\app}[1]{Appendix~\ref{#1}}

\newcommand{\sref}[1]{ref.~\citenum{#1}}

\newcommand{\eqn}[1]{Eq.~(\ref{#1})}

\newcommand{\eqnn}[2]{Eqs.~(\ref{#1}) and (\ref{#2})}

\newcommand{\no}{\nonumber}

\newcommand{\gtnote}[1]{\textcolor{red}{#1}}
\newcommand{\gtsout}[1]{\textcolor{red}{\sout{#1}}}
\newcommand{\gtmsout}[1]{\ensuremath{\textcolor{red}{\cancel{#1}}}}

\renewcommand{\gtnote}[1]{\textcolor{black}{#1}}
\renewcommand{\gtsout}[1]{}
\renewcommand{\gtmsout}[1]{}

\begin{document}

\title{Path-integral dynamics \gtnote{of water} using curvilinear centroids}

\author{George Trenins}\affiliation{ 
Department of Chemistry, University of Cambridge, Lensfield Road,\\ Cambridge,
CB2 1EW, UK.%
}%

\author{Michael J.\ Willatt}
\affiliation{ 
Laboratory of Computational Science and Modeling, IMX,
\'Ecole Polytechnique F\'ed\'erale de Lausanne, 1015 Lausanne, Switzerland.%
}%
\author{Stuart C.\ Althorpe}%
\email{sca10@cam.ac.uk}
\affiliation{ 
Department of Chemistry, University of Cambridge, Lensfield Road,\\ Cambridge,
CB2 1EW, UK.%
}%

\date{\today}

\begin{abstract}
We develop a path-integral dynamics method for water that resembles centroid
molecular dynamics (CMD), except that the centroids are averages of
curvilinear, rather than cartesian, bead coordinates. The curvilinear
coordinates are used explicitly only when computing the potential of
mean force, the components of which are re-expressed in terms of
cartesian `quasi-centroids' (so-called because they are close to the
cartesian centroids). Cartesian equations of motion are obtained by
making small approximations to the quantum Boltzmann distribution.
Simulations of the infrared spectra of various water models over
150--600~K show these approximations to be justified: for a
two-dimensional OH-bond model, the quasi-centroid molecular dynamics
(QCMD) spectra lie close to the exact quantum spectra, and almost on
top of the Matsubara dynamics spectra; for gas-phase water, the QCMD
spectra are close to the exact quantum spectra; for liquid water and
ice (using the q-TIP4P/F surface), the QCMD spectra are close to the
CMD spectra at 600~K, and line up with the results of thermostatted
ring-polymer molecular dynamics and approximate quantum calculations at
300 and 150~K. The QCMD spectra show no sign of the CMD `curvature
problem' (of erroneous red shifts and broadening). In the liquid and
ice simulations, the potential of mean force was evaluated on-the-fly
by generalising an adiabatic CMD algorithm to curvilinear coordinates;
the full limit of adiabatic separation needed to be taken, which made
the QCMD calculations 8~times more expensive than partially adiabatic
CMD at 300~K, and 32~times at 150~K (and the
intensities may still not be converged at this temperature). The QCMD method is probably
generalisable to many other systems, provided collective
bead-coordinates can be identified that yield  compact mean-field
ring-polymer distributions.
\end{abstract}
\maketitle

\section{\label{sec:intro}Introduction}

A major challenge in molecular simulation is to include nuclear quantum effects
in the dynamics of liquid water, ice, and related systems.  One way to do this
is to make local approximations around one water molecule, then to solve the
Schr\"odinger equation in a reduced
space.\cite{skinner2009,skinner2013,lmon1,lmon2} Another way, arguably more
general, because it does not require a local approximation, is to use
path-integrals to describe the quantum Boltzmann statistics exactly, and
classical mechanics to approximate the dynamics.  \cite{billjpc,liubill,liurev,shig,jens1,
xantheas,liuwat,bonella,vollume,van1,van2,jerin1,jerin2,caov2,craig1,craig2,tfmiller1,annurev,trpmd1,jorsca,tim,tom1,tom2,paes1,paes2} This type of
approach assumes that real-time coherences mainly average out in
condensed-phase systems, leaving most of the quantum effects in the statistics.

The oldest such `quantum statistics + classical dynamics' approach is based on
the classical Wigner, or LSC-IVR,
approximation.\cite{billjpc,liubill,liurev,shig,jens1} This approach is
powerful, and has been used to calculate infrared\cite{xantheas} and recently
Raman spectra\cite{liuwat} for liquid water. However, LSC-IVR has the drawback
that the classical trajectories do not conserve the quantum Boltzmann
distribution, which has the effect of damping spectral lineshapes. 

More recently, approaches such as centroid molecular dynamics (CMD)\cite{caov2}
and (thermostatted) ring-polymer molecular dynamics
((T)RPMD)\cite{craig1,craig2,tfmiller1,annurev,trpmd1,jorsca,tim,tom1,tom2} have been developed,
which redress this problem by using trajectories that do conserve the quantum
Boltzmann distribution. These approaches are very practical, since they can be
run as extended classical MD simulations. But they also have drawbacks of their
own.

The CMD method has proved \gtnote{to be useful for} simulating water dynamics, most
notably in recent calculations on the MB-pol {\em ab initio} water potential
energy surface.\cite{paes1,paes2} The method works by generating classical
trajectories on the free-energy surface obtained by mean-field averaging the
quantum Boltzmann distribution around the centroids (centres-of-mass) of the
ring-polymers (imaginary-time Feynman paths\cite{chanwol,pariram,ceperley}). 
However, CMD breaks down at low temperatures, when the mean-field ring-polymer
distributions delocalise.\cite{trpmd2,marx09,marx10} 
When applied to gas-phase water, 
the CMD ring-polymer distribution spreads in a crescent around the curve of the
rotating OH-bond, so that the centroid lies approximately at its focal point,
which flattens the potential, and leads to an artificially red-shifted and
broadened spectral lineshape. This `curvature problem' is less severe in the
condensed than the gas phase,\cite{marx10} since there is less spreading around librations
than rotations, but it is still serious enough to give, e.g., a 100 cm$^{-1}$
red shift and considerable broadening to the OH-stretch band of ice at 150~K.\cite{trpmd2}

The RPMD method does not suffer from the CMD curvature problem, since it
involves no mean-field averaging, propagating instead all the ring-polymer
`beads' (imaginary-time replicas) as separate particles. However, the dynamics
of the modes orthogonal to the centroid are fictitious, because of the
ring-polymer spring-forces, and hence need to be damped by a thermostat to
avoid corrupting the spectrum.\cite{trpmd1,trpmd2,trpmd3} This means that
thermostatted (T)RPMD is an extremely efficient method for generating spectral
line positions, but that it predicts lineshapes that are often too broad. 

It would therefore be useful to develop a path-integral method which generates
dynamics that is quantum Boltzmann conserving, but which manages to avoid both
the curvature problem of CMD, and the fictitious springs of RPMD. In refs.~\onlinecite{marx09,marx10},
Marx and co-workers suggest that such a method could be developed  by modifying CMD,
such that the centroids are taken as the averages of
curvilinear rather than cartesian bead coordinates. This
curvilinear centroid would lie at the centre of the ring-polymer distribution,
rather than at its approximate focal point, which would stop the free energy
from artificially flattening. To our knowledge, this
suggestion has not yet been pursued.
 
More recently, the case for curvilinear centroids has been strengthened
further, by the development of `Matsubara dynamics', which is the
quantum-Boltzmann-conserving classical dynamics that results when the exact
quantum dynamics is mean-field averaged to make it smooth and continuous in
imaginary time.\cite{mats,mf-mats} Matsubara dynamics is impossible to compute
exactly (except in toy systems), owing to a phase problem, but it can be
computed approximately.\cite{jens2,planets,mats-rpmd} In fact CMD and TRPMD
(both first obtained heuristically) are
mean-field and short-time approximations to Matsubara dynamics.\cite{mats-rpmd}
By considering a two-dimensional model, it was shown that the CMD
curvature problem in the OH stretch can be cured if just a few modes are
released from the centroid-averaged mean-field.\cite{mf-mats}  The resulting Matsubara
stretch bands are very close to the exact quantum results, the only significant
error being a small, temperature-independent blue shift, which also affects CMD
(when the temperature is high enough to avoid the curvature problem) and TRPMD.
It was also found that the  spreading of the CMD ring-polymer distribution
around the rotational angle is exacerbated by artificial instantons, which form
at small bond-lengths because of the cartesian centroid constraint.\cite{mf-mats} Curvilinear
centroids would avoid such artefacts, and should therefore give more compact
distributions around the rotational angle, and thus a better approximation to
the Matsubara-dynamics spectrum.

In this article, we investigate these likely advantages, by developing a method,
called `quasi-centroid molecular dynamics' (QCMD), which applies curvilinear
centroid dynamics to water. We have not tried formally to derive QCMD from
Matsubara dynamics---such a derivation is almost certainly possible, but we
thought it best first to test QCMD numerically.

Before going further, however, we mention some of the concerns that have
probably put people off using curvilinear centroids to date.  First,
curvilinear hamiltonians are notoriously difficult to derive and work
with,\cite{miller-ham,tennyson-ham,truhlar-ham} especially when used for
path-integrals (which exhibit strange pathologies when formulated in
curvilinear coordinates\cite{kleinert,marx-curv}). Second, in using
non-cartesian centroids, one runs the risk that components of the ring-polymer
springs could survive the mean-field averaging, and thus corrupt the forces.
Third, except in the high-temperature limit, the overall curvilinear centroid
(which we will refer to as the `quasi-centroid') does not coincide with the
overall cartesian centroid (which lies at the bead centre-of-mass); as a
result,  the quasi-centroid  does not give the exact estimator for a linear
operator. Our response to these concerns is to
formulate the mean-field dynamics in as cartesian a way as possible, by making
approximations to the quantum Boltzmann distribution  which are expected to be
minor, provided it is sufficiently compact that
the quasi-centroids are close to the centroids.  The numerical results in
\ssec{sec:2d-morse}, \ref{sec:ps-h2o}, and \ref{sec:application} will show that
these assumptions appear to be justified.

The article is structured as follows. \lsec{sec:2d-morse} develops the QCMD method for a
two-dimensional OH-stretch model, for which the curvilinear centroid is simply
the average of the bead radii. The resulting QCMD spectra are found to lie
close to the exact quantum spectra, and almost on top of the Matsubara spectra.
\lsec{sec:ps-h2o} extends these results to gas-phase water, by using bond-angle
centroids, obtaining similarly close agreement with the exact quantum results.
\lsec{sec:aqcmd} shows how to compute the mean-field quasi-centroid forces
on-the-fly, using a modification of the adiabatic CMD (ACMD)
algorithm,\cite{caov4,hone1,hone2} with SHAKE and
RATTLE\cite{ryckaerts,andersen,tuckerman} used to impose curvilinear
constraints.  \lsec{sec:application} reports applications of QCMD to liquid
water and ice, using the bond-angle centroids of \ssec{sec:ps-h2o} to describe
the internal motion of the monomers, and an
Eckart-like rotational frame\cite{eckart,bright-wilson} 
to describe their translational and librational motion.
\lsec{sec:conclusions} concludes the article with suggestions of how to
generalise QCMD to systems other than water.

\section{Curvilinear centroids in a two-dimensional model%
\label{sec:2d-morse}}

To develop the key ideas behind QCMD, we first consider a familiar two-dimensional `champagne-bottle' model
(details given in \app{app:spectra-calc}) of a vibrating-rotating OH bond, with a linear
dipole moment operator.

\subsection{Standard CMD treatment}
In CMD, one considers the motion of the centroid, which is the centre of mass of the ring-polymer.\cite{caov2} For a particle of mass $m$ moving in two dimensions, represented by $N$ ring-polymer beads, the centroid position ${\bf Q}\equiv(Q_{x},Q_{y})$ is 
\begin{align}
{\bf Q}=\dfrac{1}{N}\sum_{i=1}^N {\bf q}_{i}
\end{align}
where ${\bf q}_i\equiv(q_{ix},q_{iy})$ are the cartesian coordinates of the $i$-th ring-polymer bead. 
The CMD equations of motion are
\begin{align}
\dot {\bf Q} =& {\dfrac{\bf P}{m}}\no\\
\dot {\bf P}=&-{\dfrac{\partial F({\bf Q})}{\partial {\bf Q}}}
\end{align}
where ${\bf P}$ is the centroid momentum, and
\begin{align}
F({\bf Q})=&-\dfrac{1}{\beta}\ln Z_0({\bf Q})
\end{align}
is the free energy obtained from\cite{reduced}
\begin{align}
Z_0({\bf Q})= &\int\! d{\bf q}'\,e^{-\beta W({\bf q}')}
 \prod_{\mathclap{\nu=x,y}}\delta\left(Q_\nu'-Q_\nu\right)
\end{align}
in which
\begin{align}
W({\bf q}) & = U({\bf q}) + S({\bf q}) \label{eq:rp-pot}\\
U({\bf q}) & = \dfrac{1}{N}\sum_{i=1}^NV({\bf q}_i)\\
S({\bf q}) & =  \dfrac{m N}{2(\beta\hbar)^2} \sum_{\nu=x,y}\sum_{i=1}^N (q_{\nu\, i+1}-q_{\nu i})^2 %
\label{eq:rp-springs} 
\end{align}
Since the spring potential $S({\bf q}) $ is independent of ${\bf Q}$,  it follows that
\begin{align}
-{\dfrac{\partial F({\bf Q})}{\partial {\bf Q}}}=-\left\langle 
\dfrac{\partial U({\bf q})}{\partial{\bf Q}}
\right\rangle_{\bf Q}
\end{align}
where
\begin{align}
\left\langle \ldots \right\rangle_{\bf Q} & = \dfrac{1}{Z_0({\bf Q})}\int\! d{\bf q}'\,e^{-\beta W({\bf q}')}
 (\ldots)\,\prod_{\mathclap{\nu=x,y}}\delta\left(Q_\nu'-Q_\nu\right)
\end{align}
which is useful in practical calculations, since the right-hand side can be
evaluated using methods such as adiabatic centroid molecular dynamics (ACMD),\cite{caov4,hone1,hone2}
as discussed in \ssec{sec:aqcmd}.

The CMD approximation to the infrared spectrum is obtained from the Fourier transform of
\begin{align}
C(t)=& \dfrac{1}{(2\pi\hbar)^2 Z}\int\! d{\bf P}\int\! d{\bf Q}\,e^{-\beta H({\bf P},{\bf Q})}\dot{\bm \mu}({\bf Q})\!\cdot\!\dot{\bm \mu}({\bf Q}(t))
\label{eq:2d-cmd-tcf}
\end{align}
with
\begin{align}
H({\bf P},{\bf Q}) = \dfrac{\mathbf{P}^2}{2m}+F({\bf Q})
\end{align}
In the model of \app{app:spectra-calc}, we use a linear dipole moment $\bm{\mu} =
\mathbf{Q}$, and multiply $C(t)$ by a time-window function before taking the
Fourier transform.

\fig{fig:2d-spectra} illustrates the performance of CMD for the two-dimensional OH-bond
model. As is well known, CMD works well for this model at relatively high
temperatures, but the `curvature problem' artificially red-shifts and broadens
the OH stretch at lower temperatures,  where the centroid-constrained ring
polymers spread around the potential curve.\cite{marx09,marx10}  In \sref{mf-mats}, it was shown that this
behaviour  results from artificial instantons minimising the ring-polymer potential energy
at small bond-lengths. The curvature problem  worsens noticeably once the
temperature is low enough for the CMD trajectories to encounter these
instantons, as happens at about $200~\si{K}$ in the OH model (see
\sfig{fig:2d-spectra} showing the spectra and \sfig{fig:2d-blobs} showing an
instanton-forming CMD trajectory). That the instantons minimise the
ring-polymer energy is an artefact of applying a cartesian constraint to a
curved potential: by stretching the polymers round the curve, one can push them
outwards (away from the repulsive wall) without affecting the constraint.

\subsection{Quasi-centroid molecular dynamics}
\label{qcmd2d}

We therefore propose constraining the centroid
\begin{align}
R & = \dfrac{1}{N}\sum_i r_i
\end{align}
of the ring-polymer radial coordinates
\begin{align}
r_i & = \sqrt{x_i^2+y_i^2}
\end{align}
This constraint makes it impossible for the polymers  to lower their energy by
stretching and moving outwards into instantons, and ensures that $R$ describes
the centre of the ring-polymer distribution rather than its approximate focal
point.\cite{instantons} We also define an analogous polar angle $\Theta$, which runs from $0$ to $2\pi$ such that the cartesian coordinates
\begin{align}
{\overline Q}_x & = R \cos \Theta\no\\
{\overline Q}_y & = R \sin \Theta
\end{align}
cover the whole two-dimensional space.  Note that the dependence of $\Theta$ on the bead coordinates need not be specified, because the circular symmetry of the model means that the \gtnote{potential of mean force} is independent of~$\Theta$.

The non-linear relation between the cartesian and polar bead coordinates means that
$\overline {\bf Q}\ne{\bf Q}$. However, it is easy to show (by expanding $R$ in
terms of ring-polymer normal modes) that $\overline {\bf
Q}\simeq{\bf Q}$ if the distribution is reasonably compact, and that $\overline
{\bf Q}\to{\bf Q}$ in the high temperature limit.  We will therefore refer to
$\overline {\bf Q}$ as the position of the `quasi-centroid'.

In quasi-centroid molecular dynamics (QCMD), we use cartesian equations of
motion which resemble those of CMD, except that the mean-field averages are
taken around the quasi-centroid rather than the centroid:
\begin{subequations}\label{both}
\begin{align}
\dot{\overline {\bf Q}} & = \dfrac{\overline{\mathbf{P}}}{m}%
    \label{eq:2dp}\\
\dot{\overline {\bf P}} & =-\dfrac{\partial \overline F(R)}{\partial \overline {\bf Q}}%
    \label{eq:2df}
\end{align}
\end{subequations}
where
\begin{align}
\overline F(R) & = -\dfrac{1}{\beta}\ln \dfrac{{\overline Z}_0(R)}{R}
\label{eq:2d-free-energy} 
\end{align}
with
\begin{align}
{\overline Z}_0(R) & =  \int d{\bf q}' e^{-\beta W({\bf q}')} \,\delta\!\left(R'-R\right),
\end{align}
In writing out these equations we have assumed that it is a good approximation
to mean-field average the exact quantum dynamics about the quasi-centroid phase
space $(\overline{\bf P},\overline{\bf Q})$, and that this mean-field-averaged
dynamics is classical.  The first assumption is expected to hold if the
distribution is compact, and the second one if the mean-field-averaging gives a
phaseless Matsubara dynamics (which we have not derived, but which seems
likely).  In writing \eqn{eq:2df}, we have also assumed that $\overline{\bf P}$
can be approximated to be {\em purely cartesian}, such that it  contributes a
factor of $\exp(-\beta\overline{\bf P}^2\!/2m)$ to the quantum Boltzmann
distribution.
\app{app:qc-approx} shows that this last assumption can be expected to hold if the
ring-polymer distribution is sufficiently compact.

\begin{figure}
\includegraphics{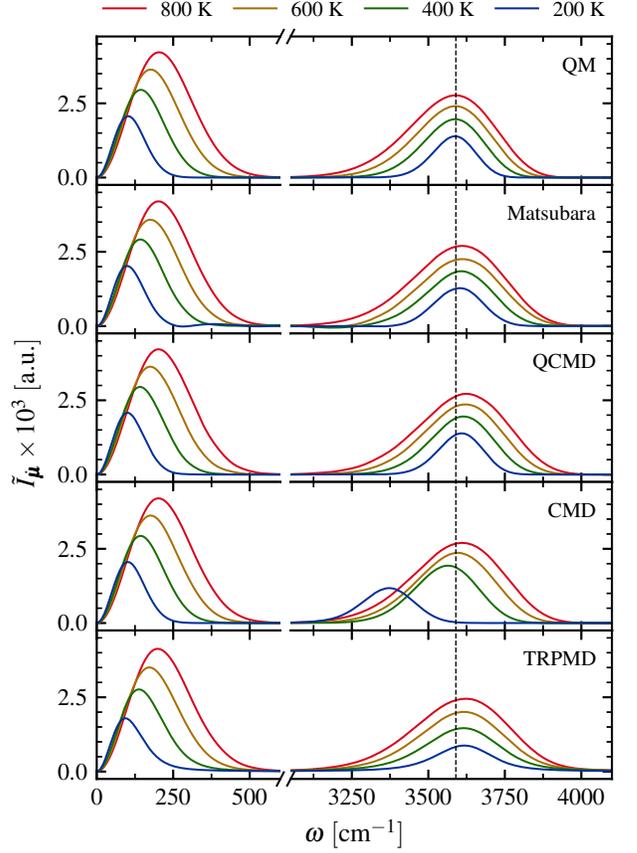}
\caption{\label{fig:2d-spectra} 
Simulated absorption spectra for the two-dimensional OH-bond model. The
curvilinear centroids used in QCMD yield  good agreement with the exact
quantum (QM) and Matsubara-dynamics spectra, whereas the cartesian
centroids used in CMD  artificially red-shift and broaden  the
OH-stretch bands.  The QCMD results were calculated as described in
\ssec{sec:2d-morse}; all other results are taken from \sref{mf-mats}.
     }
\end{figure}

We further simplify the calculation of the force by making the approximation
\begin{align}\label{eq:fap}
-\dfrac{\partial\overline{F}(R)}{\partial \overline{\bf Q}} \simeq
-\left \langle \dfrac{\partial U(\mathbf{q})}{\partial \overline{\bf Q}} \right\rangle_R
\end{align}
where
\begin{align}
\left \langle\dots  \right\rangle_R& = \dfrac{1}{{\overline Z}_0(R)}  \int d{\bf q}'  e^{-\beta W({\bf q}')} (\ldots)\,\delta\!\left(R'-R\right),
\end{align}
which is equivalent to assuming that the polymer-spring forces do not survive
the mean-field averaging.\cite{Jacob} Unlike its counterpart in CMD,
\eqn{eq:fap} is not exact, since $R$ contains components of the ring-polymer
normal modes orthogonal to the centroid. However, the size of this spring-force
will depend on the extent to which the ring-polymer distribution spreads and
contracts as a function of $R$, and is thus expected to be small if the
distribution remains compact. The advantage of \eqn{eq:fap} is that
it will allow us later (\ssec{sec:aqcmd}) to evaluate the force on-the-fly
using a simple algorithm.
The cartesian forces of \eqn{eq:fap} are obtained from the forces on $R$ using 
\begin{flalign}
- \dfrac{\partial U(\mathbf{q})}{\partial \overline{\bf Q}}  & = \dfrac{\overline{\bf Q}}{R}  f_{R}(\mathbf{q})  \label{eq:2d-mf-force}
\end{flalign}
with
\begin{flalign}
f_R(\mathbf{q}) & \equiv -\dfrac{\partial U(\mathbf{q})}{\partial R} = -\dfrac{1}{N} \sum_{i = 1}^{N} \dfrac{\mathbf{q}_{i}}{r_i} \! \cdot \! \bm{\nabla} V(\mathbf{q}_i).%
\label{eq:2d-qcmd-fr}
\end{flalign}

Having propagated the quasi-centroid trajectory $\overline Q(t)$, the spectrum
is then obtained from the time-correlation function
\begin{align}
\overline C(t) & = 
\dfrac{1}{(2\pi\hbar)^2\overline{Z}}
\int\! d\overline{\mathbf{P}}\int\!d\overline{\bf Q}\, 
  e^{-\beta \overline {H}(\overline{\bf P},\overline{\bf Q})}\, 
  \dot{\bm{\mu}}(\overline{\mathbf{Q}})\!\cdot\!\dot{\bm{\mu}}(\overline{\mathbf{Q}}(t))%
\label{eq:2d-qcmd-tcf}
\end{align}
with
\begin{align}
\overline Z & = \dfrac{1}{(2\pi\hbar)^2}
\int\! d\overline{\mathbf{P}} \int\! d\overline{\mathbf{Q}}\,
e^{-\beta \overline{H}(\overline{\mathbf{P}},\overline{\mathbf{Q}})}
\end{align}
and
\begin{align}
\overline{H}(\overline{\bf P},\overline{\bf Q})=\overline{\bf P}^{2\!}/2m + f(\overline{\mathbf{Q}})
\end{align}
where $f(\overline{\mathbf{Q}})\simeq \overline F(\overline{\mathbf{Q}})$ is
defined to be the free energy obtained by doing work with the approximate force
of \eqn{eq:fap}.

In a QCMD simulation, one propagates trajectories using \eqnn{both}{eq:fap},
applying a standard thermostat in order to sample the distribution
$\exp(-\beta\overline {H}(\overline{\bf P},\overline{\bf Q}))$. As already
mentioned, this distribution involves approximations to the exact quantum
Boltzmann distribution which would not have been made in the analogous CMD
calculation. In addition, CMD has the advantage of using the exact estimator
for the linear dipole moment operator ${\bm \mu}={\bf Q}$, whereas QCMD uses
the approximation ${\bm \mu}\simeq\overline{\bf Q}$. We thus
expect QCMD to be less accurate than CMD when the temperature is high enough to
neglect the curvature problem, but more accurate at lower temperatures, where
the better treatment by QCMD of the dynamics will far outweigh the errors in the
Boltzmann distribution which are expected to be small (provided the
ring-polymer distribution is compact---see \app{app:qc-approx}). To be safe, we can
monitor the errors in the distribution numerically, by comparing QCMD
static properties with results from standard \gtnote{path-integral molecular dynamics
(PIMD)} calculations.

We tested QCMD on the two-dimensional OH-bond model \gtnote{of \app{app:spectra-calc}}. The forces
in~\eqn{eq:2d-qcmd-fr} were evaluated at 128 grid-points along the $R$
coordinate, using standard PIMD\cite{tuckerman} with the quasi-centroid constraint
imposed using \mbox{SHAKE} and
\mbox{RATTLE}.\cite{ryckaerts,andersen,tuckerman} The resulting spectra
(\sfig{fig:2d-spectra}) are in excellent agreement with the exact quantum and
Matsubara results, at all temperatures tested, with no sign of a curvature
problem. \fig{fig:2d-qcmd-mats} makes a closer comparison of the QCMD
OH-stretch bands with the mean-field Matsubara results of \sref{mf-mats}. We
see that
the QCMD OH-stretch peaks have small ({$\sim\!10~\mathrm{cm}^{-1}$) blue shifts
with respect to the Matsubara peaks. The shifts change slightly with
temperature, ranging from $14~\mathrm{cm}^{-1}$ at $800~\mathrm{K}$ to
$6~\mathrm{cm}^{-1}$ at $200~\mathrm{K}$. This variation appears to be the
result of small differences in lineshape, rather than an underlying curvature
problem; part of these differences may be the result of convergence artefacts
in the Matsubara results. 
\begin{figure}
\includegraphics{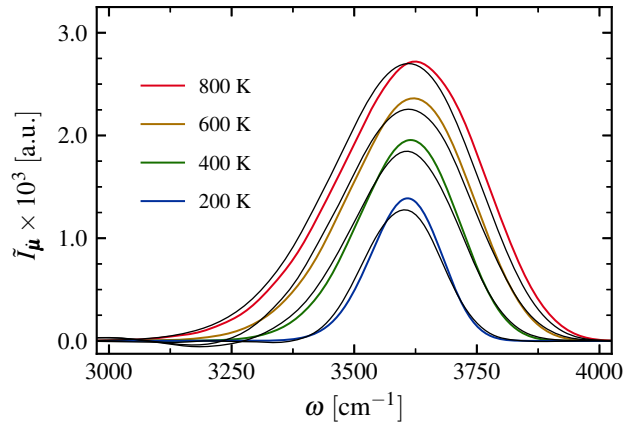}
\caption{\label{fig:2d-qcmd-mats} 
A closer look at the QCMD (coloured lines) and Matsubara (black lines)
spectra of \sfig{fig:2d-spectra} in the region of the OH stretch band.}
\end{figure}

\begin{figure}
\includegraphics{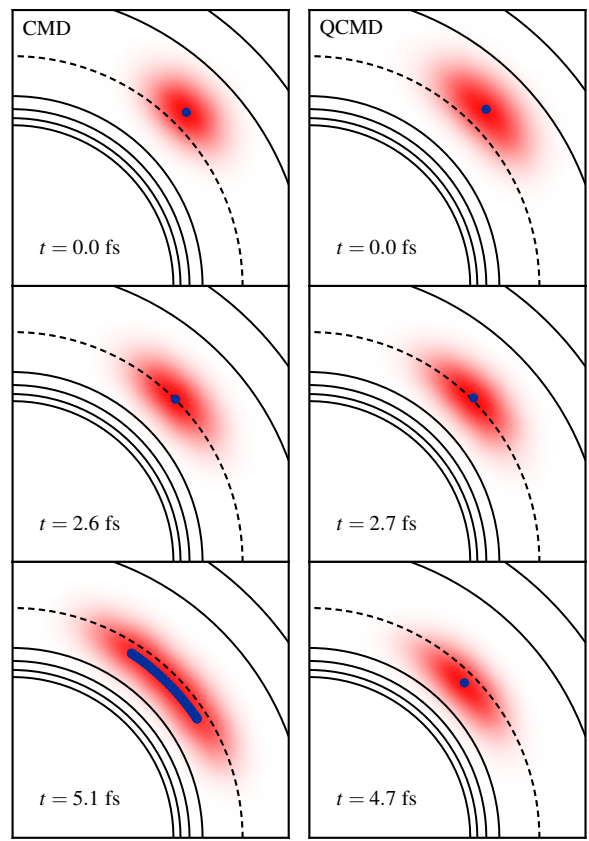}
\caption{\label{fig:2d-blobs} 
        Snapshots of the CMD and QCMD ring-polymer bead distributions (red),
        along a pair of trajectories with extremely short (but thermally
        accessible) inner turning points, taken from the 200~K OH-model
        simulations. At the inner turning point (bottom),  the CMD distribution
        fluctuates around an artificial instanton (blue blobs), whereas the
        QCMD distribution fluctuates around a point. The dashed contour
        indicates the equilibrium bond-length $r_{\mathrm{eq}}$, and the solid
        black contours are drawn at energy increments of~$0.2 D_0$.
        }
\end{figure}

As expected, the QCMD distributions contain no artificial instantons at the
inner turning points (see \sfig{fig:2d-blobs}).  Detailed comparison
with CMD trajectories is difficult (since the dynamics are different), but it
appears that the QCMD distributions are broader than the CMD
distributions at the outer turning points. This property may perhaps be
responsible for the small blue shifts in comparison with the Matsubara results
(see \sfig{fig:2d-qcmd-mats})---although if this were case, one would expect a
systematic
temperature dependence. Overall, the widths of the QCMD distributions vary little over a vibrational period, confirming that 
dynamics of the quasi-centroid is coupled only weakly to the internal degrees of freedom, so that the neglect of 
the ring-polymer springs in \eqn{eq:fap} is justified.

Given the excellent agreement of the QCMD and Matsubara spectra, we \gtnote{expect} that
the various QCMD static approximations mentioned above are small.  This
is confirmed by comparing QCMD static averages with the results of exact
quantum calculations. We found that the QCMD expectation values of the
bond-length $R$ were identical to within sampling error to the exact values, at
all temperatures tested, and that the largest error in the $t=0$ limit of the
dipole autocorrelation function was 2$\%$ at 200 K. \gtnote{Note that this error 
does not affect the calculated spectra, since these were obtained from the dipole-\emph{derivative}
autocorrelation function, for which QCMD  gives the  exact $ t = 0 $ limit for a linear dipole.}

\gtnote{%
We attempted to `break' the QCMD calculations by pushing the temperature down to 50 K. However, even at this low temperature, the ring-polymer distribution around the centroid remained compact, and the agreement of QCMD with the exact quantum spectrum was as close as that shown at higher temperatures in \sfig{fig:2d-spectra}. This suggests that QCMD continues to gives a good approximation to Matsubara dynamics down to 50~K (where we were unable to compute the Matsubara spectrum because the phase problem was too severe).}

In summary, QCMD behaves extremely well for the two-dimensional OH model. It
eliminates the curvature problem, giving spectra that lie almost on top of the
Matsubara results at all temperatures tested.  The errors from the `extra'
approximations (that $\overline{\bf P}$ is cartesian and that the polymer
springs can be neglected in the force) are found to be insignificant.

\section{Extension to gas-phase water%
\label{sec:ps-h2o}}

We now test whether these advantages continue when QCMD is extended to treat
gas-phase water.  By analogy with \ssec{sec:2d-morse}, we choose a
set of curvilinear centroid coordinates that can be expected to give compact
mean-field distributions, namely the bond-angle centroids
\begin{align}
 R_1 & = \dfrac{1}{N} \sum_{i} r_{i1}, & r_{i1} & = |\mathbf{q}^{%
\scriptscriptstyle{(\mathrm{OH}_1)}%
}_{i}| %
\no \\
 R_2 & = \dfrac{1}{N} \sum_{i} r_{i2}, & r_{i2} & = |\mathbf{q}^{%
\scriptscriptstyle{(\mathrm{OH}_2)}%
}_{i}| %
 \no \\
 \Theta & = \dfrac{1}{N} \sum_{i} \theta_i, & \theta_i & = \arccos\!\left[
\dfrac{\mathbf{q}^{%
\scriptscriptstyle{(\mathrm{OH}_1)}%
}_{i} \! \cdot \! \mathbf{q}^{%
\scriptscriptstyle{(\mathrm{OH}_2)}%
}_{i}}%
{r_{i1}r_{i2}}
\right] %
\label{eq:h2o-theta}
\end{align}
in which $\mathbf{q}^{%
\scriptscriptstyle{(\mathrm{OH}_{1,2})}%
}_{i}={\bf q}^{\scriptscriptstyle{(\mathrm{H}_{1,2})}}_i\!-{\bf
q}^{\scriptscriptstyle{(\mathrm{O})}}_i$\! are the displacement vectors between
the oxygen and hydrogen atoms in the $i$-th water molecule replica ($i = 1
\ldots N$). A further six coordinates would be required to orient the
mean-field-averaged water molecule and locate its
centre-of-mass, but, like $\Theta$ in \ssec{sec:2d-morse}, these external coordinates
need not be specified, since no force or torque acts on them.

As in \ssec{sec:2d-morse}, we carry out dynamics using equations of motion
which have the same form as CMD, namely
\begin{align}
\dot{\overline {\bf Q}}^{(\alpha)} & = \dfrac{\overline{\bf P}^{(\alpha)}}{m_\alpha}\no\\
\dot {\overline {\bf P}}^{(\alpha)}& = -\dfrac{\partial\overline{F}(\bm{\xi})}{\partial \overline {\bf Q}^{(\alpha)}}
\label{eq:watermol-ham-eq}
\end{align}
where ${\overline {\bf Q}}^{(\alpha)}$  locate the quasi-centroids of atoms
\mbox{$\alpha = ({\rm H}_1,{\rm H}_2,{\rm O})$} in the mean-field averaged
water molecule. As in \ssec{sec:2d-morse}, the curvilinear coordinates enter
only through the force, which, by analogy with \eqn{eq:fap}, is approximated as 
\begin{align}
-\dfrac{\partial\overline{F}(\bm{\xi})}{\partial \overline{\bf Q}^{(\alpha)}} \simeq
-\left \langle \dfrac{\partial U(\mathbf{q})}{\partial \overline{\bf Q}^{(\alpha)}} \right\rangle_{\mathrlap{\!\!\bm{\xi}}}.%
\label{eq:watermol-fap}
\end{align}
where
\begin{flalign}
\left\langle \ldots \right\rangle_{\bm \xi} & = \dfrac{1}{\overline{z}_0({\bf \xi})}\int d \mathbf{q}' e^{-\beta W(\mathbf{q}')} (\ldots)\,
\prod_{\mathclap{\nu=1}}^{3} \delta\left(\xi_{\nu}' - \xi_{\nu}\right)\no\\
{\overline z}_0({\bm \xi}) & = \int d \mathbf{q}' e^{-\beta W(\mathbf{q}')}\,
\prod_{\mathclap{\nu=1}}^{3} \delta\left(\xi_{\nu}' - \xi_{\nu}\right),
\end{flalign}
with $\bm{\xi} \equiv \left(\xi_1,\xi_2,\xi_3 \right)\equiv \left(R_1, R_2 ,\Theta \right)$, and
${\bf q}\equiv \left\{{\bf q}_i^{(\alpha)}\right\}$. 

The right-hand side of \eqn{eq:watermol-fap} is computed using
\begin{flalign}
\hspace*{-1em}-\dfrac{\partial U(\mathbf{q})}{\partial\overline{\bf Q}^{%
\scriptscriptstyle{(\mathrm{H}_{k})}%
}}
 = {}  & \dfrac{\overline{\bf Q}^{%
\scriptscriptstyle{(\mathrm{OH}_k)}%
}}{R_k}  f_{\!R_k}\!(\mathbf{q}) - {} \no \\ 
{}  \dfrac{1}{R_k \sin \Theta} & \left(
\dfrac{\overline{\bf Q}^{%
\scriptscriptstyle{(\mathrm{OH}_j)}%
}}{R_j} - \dfrac{\overline{\bf Q}^{%
\scriptscriptstyle{(\mathrm{OH}_k)}%
}}{R_k} \cos \Theta
\right)\!  f_{\Theta}(\mathbf{q}) \no\\
\vphantom{\dfrac{\overline{Q}^{%
\scriptscriptstyle{(\mathrm{OH}_k)}%
}_{\nu}}{R_k}}%
- \dfrac{\partial U(\mathbf{q})}{\partial\overline{\bf Q}^{%
\scriptscriptstyle{(\mathrm{O})}%
}} =  {} &
\sum_{k = 1,2}\dfrac{\partial U(\mathbf{q})}{\partial\overline{\bf Q}^{%
\scriptscriptstyle{(\mathrm{H}_k)}%
}}, \label{eq:qcmd-oforce}
\end{flalign}
with $(k,j)=(1,2)$ or $(2,1)$, and
\begin{flalign}
f_{R_k}\!(\mathbf{q}) &\equiv -\dfrac{\partial U(\mathbf{q})}{\partial R_k}  = -\dfrac{1}{N} \sum_{i=1}^{N}\dfrac{\mathbf{q}^{%
\scriptscriptstyle{(\mathrm{OH}_k)}%
}_{i}}{r_{ik}} \cdot\! \bm{\nabla}_{\!\mathrm{H}_k\!} V(\mathbf{q}_i)\no\\ 
 f_{\Theta}(\mathbf{q}) \equiv & -\dfrac{\partial U(\mathbf{q})}{\partial \Theta} = \dfrac{1}{N} \sum_{i=1}^{N} 
\dfrac{r_{i1}}{\sin\theta_i} \times \no \\
& \left( \dfrac{\mathbf{q}^{%
\scriptscriptstyle{(\mathrm{OH}_2)}%
}_{i}}{r_{i2}} - 
\dfrac{\mathbf{q}^{%
\scriptscriptstyle{(\mathrm{OH}_1)}%
}_{i}}{r_{i1}} \cos\theta_{i} 
\right) \cdot \bm{\nabla}_{\!\mathrm{H}_1\!} V(\mathbf{q}_i),%
\label{eq:dUdTheta}
\end{flalign}
where $\bm{\nabla}_{\!\mathrm{H}_k\!} V(\mathbf{q}_i)$ is the force on the
$i$-th bead of the \mbox{$k$-th} hydrogen. The infrared spectrum is obtained
from
\begin{flalign}
\overline{C}_{\mu\mu}(t) = \dfrac{1}{(2 \pi \hbar)^9 \overline Z} \int \! d\overline{\mathbf{P}} \! 
\int \! & d\overline{\mathbf{Q}} \,e^{-\beta \overline { H}(\overline{\bf P},\overline{\bf Q})} \nonumber \\
{} & \times \bm{\mu}(\overline{\mathbf{Q}}) \! \cdot \! \bm{\mu}(\overline{\mathbf{Q}}(t)),
\end{flalign}
with  
\begin{flalign}
\overline {H}(\overline{\bf P},\overline{\bf Q}) & =
\sum_{\alpha} \left[
\overline{\mathbf{P}}^{(\alpha)}
\right]^{\mathrlap{2}\,}/2 m_{\alpha}+f(\overline{\mathbf{Q}})%
\label{eq:watermol-qcmd-ham}
\end{flalign}
where $f(\overline{\mathbf{Q}})\simeq \overline F(\overline{\mathbf{Q}})$ is
the free energy obtained by doing work with the approximate force of
\eqn{eq:watermol-fap}.
\begin{figure}
\includegraphics{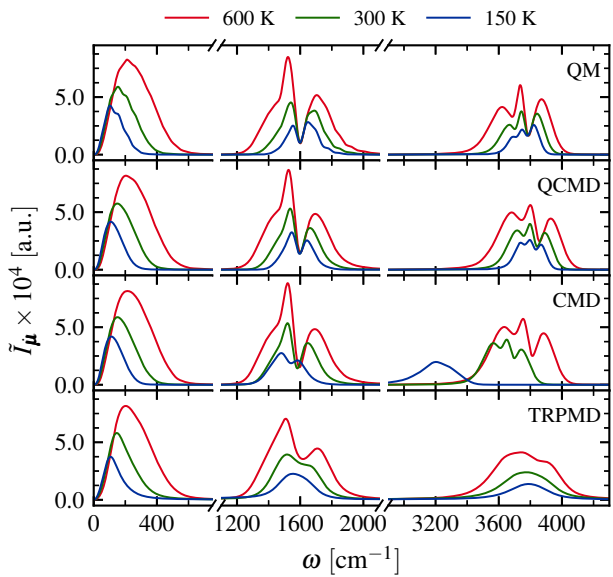}
\caption{\label{fig:ps-spectra} 
        Simulated rovibrational absorption spectra for gas-phase water,
        obtained \gtnote{after} damping the time-correlation function with a Hann window of
        width $\tau=750~\si{fs}$. The QCMD results were calculated using the
        bond-angle centroids of \ssec{sec:ps-h2o}, and show good agreement with
        the exact quantum results (QM), with no sign of the red shift and
        broadening seen in the CMD stretch.  The absorption intensities are
        scaled by a factor of 4.5 for $\omega > 1200~\mathrm{cm}^{-1}$.  
}
\end{figure}

We used the above framework to simulate QCMD infrared spectra for gas-phase
water using the  Partridge-Schwenke potential surface\cite{pspes} and dipole
moment.\cite{psdms} As in \ssec{sec:2d-morse}, we damped the time-correlation
functions before calculating their Fourier transform, this time using a Hann
window\cite{nr77} of 750~\si{fs}.
The quasi-centroid forces in \eqn{eq:dUdTheta} were calculated  on a
$64\times64\times64$ grid, using standard
PIMD\cite{tuckerman} with
\mbox{SHAKE}/\mbox{RATTLE} to impose the quasi-centroid
constraints,\cite{ryckaerts,andersen,tuckerman} and were then fitted to cubic
splines.\cite{nr77} At 600 and 300~K, the grids ranged from 1.50 to 2.50~a.u.\
in $R_{1,2}$ and from $85^{\circ}$ to $130^{\circ}$ in $\Theta$;  at 150~K this
range was reduced to 1.65 -- 2.05~a.u.\ and $90^{\circ}$\,-- $126^{\circ}$. We
also calculated the exact quantum spectrum (subject to the same damping) using
the DVR3D package of Tennyson and co-workers,\cite{dvr3d} and the corresponding
CMD and TRPMD spectra using standard PIMD techniques. A comparison with
Matsubara dynamics was not possible because the Matsubara sign problem was too
severe.

From \sfig{fig:ps-spectra}, we see that QCMD works extremely well for gas-phase
water. The overall agreement with the exact quantum results is excellent, even
at 150~K, where there is again no sign of a curvature problem. There are small
differences between the QCMD and quantum spectra:  the QCMD bend
($\sim\!1600~\mathrm{cm}^{-1}$) and stretch
\mbox{($\sim\!3800~\mathrm{cm}^{-1}$)} have different internal structures from
those in the exact spectrum, and the stretch is blue-shifted by about
$60~\mathrm{cm}^{-1}$. The main cause of these differences is most likely the
neglect of real-time coherence by QCMD.  Note that the positions of the TRPMD
bands are very close to those of the QCMD bands, implying that, as expected,
TRPMD gives a good description of the short-time dynamics of gas-phase water.

As in the two-dimensional model, the gas-phase water QCMD distributions remain
compact at inner turning points because there are no quasi-centroid-constrained
instantons. \fig{fig:ps-water-blobs} compares the distributions at inner
turning points of an `extreme'  QCMD and CMD trajectory at 150~K. As in the
two-dimensional case, the QCMD distribution remains localised whereas the CMD
distribution spreads around a centroid-constrained instanton.  This artefact
distorts the approximate axial symmetry of the distribution around the OH
bond  into a banana shape.

\begin{figure}
\includegraphics{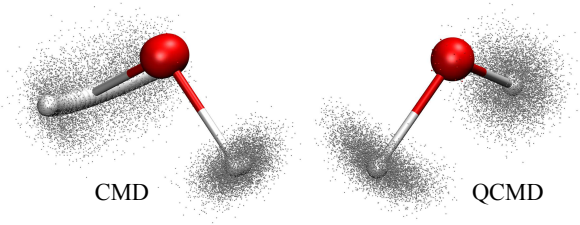}
\caption{\label{fig:ps-water-blobs} 
Ring-polymer distributions (dots) at the inner turning points of CMD and QCMD
trajectories corresponding to very short (but thermally accessible) OH
bond lengths, taken from the gas-phase water simulations at 200~K. The
CMD distribution fluctuates around an artificial instanton 
(spheres), whereas the QCMD distribution fluctuates around a localised
geometry.         }
\end{figure}

Comparisons of the QCMD average OH bond lengths with those obtained using
standard PIMD give tiny differences ($0.1\%$ at 150~K), indicating that,
as for the two-dimensional model,
the approximate distribution in~\eqn{eq:watermol-qcmd-ham} is very close to the
exact quantum Boltzmann distribution. Note that, since the Partridge-Schwenke
dipole moment
is non-linear,\cite{psdms} neither CMD nor QCMD give the exact quantum Kubo
time-correlation function in the $t \to 0$ limit,\cite{hone2,paes2008} and that
both methods neglect the Matsubara fluctuation terms necessary to completely
describe the dipole-moment dynamics at low temperatures.\cite{mats,mats-rpmd,mf-mats}

\section{Implementing QCMD using an adiabatic propagator%
\label{sec:aqcmd}}

In \ssecc{sec:2d-morse}{sec:ps-h2o}, we applied QCMD to systems that are small
enough for the mean-field force on the quasi-centroid to be pre-calculated on a
grid. To apply QCMD to larger systems, we adapt the adiabatic CMD (ACMD)
algorithm, used in previous work\cite{caov4,hone1,hone2} to calculate the mean-field centroid force on-the-fly.

In ACMD, the ring polymer is represented in terms of the normal modes that
diagonalise the ring-polymer spring potential. These
are Fourier modes, such that for each cartesian degree of freedom the lowest
frequency mode $s_0$ is (proportional to) the centroid.\cite{coalson} The mass
associated with each non-centroid mode
$s_n$ is then scaled by a factor of $( \beta \hbar \omega_n /\gamma N )^2$,
where $\omega_n$ is the associated normal-mode frequency, and a local
thermostat is applied to its dynamics.  The limit $\gamma\to\infty$ gives a
perfect adiabatic separation between the motion of the centroid and the
thermostatted normal modes, and is thus equivalent to CMD.\cite{hone2} In practice,
$\gamma$ is treated as a convergence parameter, with e.g.\ $\gamma = 64$ giving
a good approximation to CMD in gas-phase water at 150~K.

It is straightforward to modify the ACMD algorithm to an AQCMD algorithm that
adiabatically separates the motion of the quasi-centroid from that of all the
other degrees of freedom. \gtnote{%
In the AQCMD algorithm, a simulation is treated as two systems evolving in parallel, one of quasi-centroids with phase-space coordinates $ (\overline{\mathbf{P}},\overline{\mathbf{Q}}) $, and one of ring-polymers with coordinates $ (\mathbf{p},\mathbf{q}) $.  The dynamics is propagated using a modification of the integrator due to Leimkuhler and Matthews with OBABO operator splitting.\cite{leimkuhler} A propagation step begins with the thermostatting of the momenta:}
\gtnote{%
\begin{enumerate}
\item[O1.] Propagate the quasi-centroid momenta $ \overline{\mathbf{P}} $ for half a time-step ($ \Delta t /2 $) under the action of a Langevin thermostat.
\item[O2.] Propagate the bead momenta ${\bf p}$ for $ \Delta t /2 $ under the path-integral Langevin equation (PILE) thermostat,\cite{pile} followed by RATTLE\cite{andersen}  to constrain the quasi-centroid components.
\end{enumerate}
}
\gtnote{%
The momenta are then updated according to the forces acting on the system:}
\gtnote{%
\begin{enumerate}
	\item[B1.] Propagate $\overline{\bf P}$ for $ \Delta t / 2 $ under the 
	           quasi-centroid forces.
	\item[B2.] Propagate $ \mathbf{p} $ for $ \Delta t / 2 $ under the forces derived from the ring-polymer potential in~\eqn{eq:rp-pot}, followed by RATTLE.
\end{enumerate}
}
\gtnote{This is followed by a position update:}
\gtnote{%
\begin{enumerate}
	\item[A1.] Propagate the quasi-centroid positions $ \overline{\mathbf{Q}} $ for a full time-step $ \Delta t $ according to the current values of the momenta $ \overline{\mathbf{P}} $.
    \item[A2.] Propagate the bead positions $ \mathbf{q} $ for $ \Delta t $ according to the current values of $ \mathbf{p} $, followed by SHAKE,\cite{ryckaerts} which constrains the ring-polymer geometry to be consistent with the quasi-centroid configuration $ \overline{\mathbf{Q}} $ obtained at the end of step A1.
\end{enumerate}
}

\gtnote{%
The propagation step is concluded by executing B1--B2 using the forces evaluated at the updated positions, followed by O1--O2. The mass-scaling used in the propagation of the ring-polymers is the same as in ACMD, except that the mass of the centroid is also scaled, by $\gamma^{-2}$.}
A useful feature of this algorithm is that it requires explicit forces only on
the quasi-centroid (\gtnote{B1}) and on the cartesian bead coordinates (\gtnote{B2}), but does
{\em not} require explicit forces on the degrees of freedom orthogonal to the
quasi-centroid. 
Note that the AQCMD algorithm reduces to the ACMD algorithm in the special case
that the quasi-centroid is taken to be the centroid (at which point the
constraints become trivial to apply).

We tested the convergence with respect to $\gamma$ by applying the AQCMD
algorithm to gas-phase water at 150~K. A value of $\gamma = 64$ was sufficient
to reproduce the spectrum obtained by interpolating the force on a grid. The
adiabatic simulation required a time-step of $\Delta t=0.1/\gamma~\si{fs}$,
i.e.\ $\gamma$ times smaller than what would have been used to propagate the
dynamics without the mass scaling. We note that the same values of
$\gamma$ and $\Delta t$ are required to converge ACMD at this temperature, and
hence that the AQCMD algorithm uses approximately the same amount of CPU time
as the ACMD algorithm, since most of the time is spent evaluating the forces on
the ring-polymer beads. 

In condensed phase simulations, converged ACMD results can be obtained at
significantly lower adiabatic separation,\cite{hone2,mano-pacmd} provided the
strength of the thermostat acting on the non-centroid modes is carefully
tuned.\cite{trpmd2-sup} This approach is known as partially-adiabatic CMD
(PACMD). We have not yet attempted a similar partially-adiabatic implementation
of QCMD and view it as an important step in future methodological development.

\section{Liquid water and ice%
\label{sec:application}}

\gtnote{Using the AQCMD algorithm just described, we carried out} preliminary QCMD
simulations of liquid water and ice, using the q-TIP4P/F potential energy and
dipole moment surfaces.\cite{qtip4pf} We used \gtnote{sets of bond-angle centroids, defined as in}
\eqn{eq:h2o-theta} to describe the internal motion of
\gtnote{the water monomers, and `Eckart-like' rotational frames (see  \ssec{sec:eckart})
to describe the translations and overall rotations of the monomers.}
Numerical details and results of the simulations are presented in
\ssec{sec:cond-phase}.

\subsection{Eckart-like frame%
\label{sec:eckart}}
\subsubsection{Properties of the frame}

In molecular spectroscopy, the Eckart frame defines the orientation of a
molecule by means of a reference geometry, usually taken to be the equilibrium
geometry.\cite{eckart,bright-wilson} Here, we orient each water monomer using an `Eckart-like' frame, in
which the `molecule' is the monomer ring-polymer, and the `reference geometry'
is the monomer quasi-centroid. This gives the conditions
\begin{subequations}
\label{eckboth}
\begin{flalign}
  \sum_{\alpha,i} m_{\alpha} (\mathbf{q}_{i}^{(\alpha)} - \overline{\mathbf{Q}}^{(\alpha)}) & = \mathbf{0} 
  \label{eq:trans-eckart} \\
  \sum_{\alpha,i} m_{\alpha} \overline{\mathbf{Q}}^{(\alpha)} \times 
  (\mathbf{q}_{i}^{(\alpha)} - \overline{\mathbf{Q}}^{(\alpha)}) & = \mathbf{0}, %
  \label{eq:rot-eckart}
\end{flalign}
\end{subequations}
where we use the notation of \gtnote{\ssec{sec:ps-h2o}}, such that
$\mathbf{q}_{i}^{(\alpha)}$ are the atom polymer beads  and
$\overline{\mathbf{Q}}^{(\alpha)}$ are the atom quasi-centroids, with $\alpha =
({\rm H}_1,{\rm H}_2,{\rm O})$.  It is easy to show that these conditions are
equivalent to
\begin{subequations}
\label{miao}
\begin{flalign}
\sum_{\alpha} m_{\alpha} (\mathbf{Q}^{(\alpha)} - \overline{\mathbf{Q}}^{(\alpha)}) & = \mathbf{0} 
\label{eq:trans-eckart2}
\\
\sum_{\alpha} m_{\alpha} \overline{\mathbf{Q}}^{(\alpha)} \times (\mathbf{Q}^{(\alpha)} - \overline{\mathbf{Q}}^{(\alpha)}) & = \mathbf{0}.%
\label{eq:rot-eckart2}
\end{flalign}
\end{subequations}
where $\mathbf{Q}^{(\alpha)}$ are the atom centroids.

The first condition (\eqn{eq:trans-eckart} or, equivalently,
\eqn{eq:trans-eckart2}) thus places the centre-of-mass of the atom
quasi-centroids at the centre-of-mass of the entire monomer ring-polymer, or
equivalently, at the centre-of-mass of the atom centroids. 

The second condition  (\eqn{eq:rot-eckart} or, equivalently,
\eqn{eq:rot-eckart2}) specifies the orientation of the monomer ring-polymer
about its centre-of-mass. To illustrate how this works, let us imagine that we
have generated a new monomer quasi-centroid geometry $\overline{\mathbf{Q}}$
(\gtnote{using step A1} of the AQCMD algorithm), and that we now wish to define the
ensemble of ring-polymers that `belong' to this geometry. One can show that the
second Eckart condition is equivalent to minimising\cite{jorg,kudin}
\begin{flalign}
 \sum_{\alpha,i}m_\alpha|{\bf q}_{i}^{(\alpha)}- \overline{\mathbf{Q}}^{(\alpha)}|^2\no\end{flalign}
or, equivalently,
\begin{flalign}
 \sum_{\alpha}m_\alpha|{\mathbf{Q}}^{(\alpha)}- \overline{\mathbf{Q}}^{(\alpha)}|^2\no
\end{flalign}
In other words, each monomer ring-polymer in the ensemble is oriented so as to
minimise its average mass-weighted distance from $\overline{\mathbf{Q}}$. In
practice, this means that the atom centroids are usually very close to the atom
quasi-centroids---see \sfig{fig:eckart}. The
Eckart-like conditions thus ensure that the quasi-centroid-constrained
ring-polymer distribution is compact with respect to the orientational degrees
of freedom.\cite{minsp}

\subsubsection{Implementation}

To implement the Eckart-like conditions, we carry out the propagation in
cartesian coordinates $(\overline{\mathbf{P}},\overline{\mathbf{Q}})$, using
the AQCMD algorithm of \ssec{sec:aqcmd}. This allows us to avoid using
Euler angles or quaternions. Steps \gtnote{O2, B2, and A2} of the
AQCMD algorithm are straightforward to implement by imposing the
constraints of \eqn{eq:h2o-theta} and
\eqn{miao} on each monomer using SHAKE and
RATTLE.\cite{ryckaerts,andersen,tuckerman} 
Step \gtnote{B1} is implemented as follows.

First, to take into account that there are now external forces acting on the
polymer beads,  the expressions for the internal forces given in
\eqn{eq:dUdTheta} are modified according to
\begin{flalign}
\bm{\nabla}_{\!\alpha} V(\mathbf{q}_{i})  \to \bm{\nabla}_{\!\alpha} V(\mathbf{q}_{i}) + m_{\alpha}\! \left[ \dfrac{\mathbf{f}_{i}}{m_{\mathrm{tot}}} 
+ (\mathbf{I}^{-1}_{i} \bm{\tau}_{i}) \times \mathbf{q}^{(\alpha)}_{i}
\right ]\! %
\label{eq:replica-f-int}
\end{flalign}
with
\begin{flalign}
\mathbf{f}_{i} & = -\sum_{\alpha} \dfrac{\partial V(\mathbf{q})}{\partial \mathbf{q}^{(\alpha)}_{i}} 
\label{eq:eck-com-force} \\
 \bm{\tau}_{i} & = -\sum_{\alpha} \left(\mathbf{q}^{(\alpha)}_{i}-\mathbf{q}^{({\rm c})}_{i}\right) \times \dfrac{\partial V(\mathbf{q})}{\partial \mathbf{q}^{(\alpha)}_{ i}}, %
\label{eq:eck-ext-tau} 
\end{flalign}
where $\mathbf{q}_{i}^{(c)}$ is the centre-of-mass of the $i$-th replica, and
$\mathbf{I}_{i}$ is its inertia tensor in the centre-of-mass frame. Having made
this modification, we use \eqnn{eq:qcmd-oforce}{eq:dUdTheta} to compute the
{\em internal} forces on the quasi-centroid.

Second, we add the {\em external} forces on the monomer quasi-centroids, which are given by
\begin{flalign}
-\left.
\dfrac{\partial U(\mathbf{q})}{\partial \overline{\mathbf{Q}}^{(\alpha)}}
\right|_{\mathrlap{\mathrm{ext}}} =  m_{\alpha}\! \left[ \dfrac{\overline{\mathbf{f}}}{m_{\mathrm{tot}}} 
+ \left(\mathbf{I}^{-1}\overline{\bm{\tau}}\right) \times \overline{\mathbf{Q}}^{(\alpha)}
\right], %
\label{eq:qc-ext-force}
\end{flalign}
where
\begin{align}
 \overline{\mathbf{f}}&=-\sum_\alpha \dfrac{\partial U(\mathbf{q})}{\partial \overline{\mathbf{Q}}^{(\alpha)}}\no\\
 \overline{\bm{\tau}}&= -\sum_{\alpha} \left(\overline{\mathbf{Q}}^{(\alpha)}-\overline{\mathbf{Q}}^{({\rm c})}\right) \times \dfrac{\partial U(\mathbf{q})}{\partial \overline{\mathbf{Q}}^{(\alpha)}}%
 \label{eq:qc-torque-exact}
\end{align}
with $\overline{\mathbf{Q}}^{(c)}$ the quasi-centroid centre-of-mass and
$\mathbf{I}$ the  quasi-centroid inertia tensor. The centre-of-mass force
$\overline{\mathbf{f}}$ is easy to compute since it is just the average of the
forces on the replica centres-of-mass
\begin{align}
\overline{\mathbf{f}}&=\dfrac{1}{N} \! \sum_{i} \mathbf{f}_{i}
\end{align}
but the torque $\overline{\bm{\tau}}$ is difficult to compute exactly.\cite{invert}
We therefore approximate $\overline{\bm{\tau}}$
as the average of the torques on the replicas
\begin{align}
 \overline{\bm{\tau}}\simeq \dfrac{1}{N}\sum_i {\bm \tau}_i%
  \label{eq:qc-torque-approx}
\end{align}
This expression is consistent with the other approximations made to the forces
(namely that  $\overline {\bf P}$ is purely cartesian, and that
the polymer springs can be neglected in the mean-field force), since
\eqnn{eq:qc-torque-exact}{eq:qc-torque-approx} agree to within second order in
the displacements from the quasi-centroid. As with these other approximations,
the reliability of \eqn{eq:qc-torque-approx} can be monitored by comparing QCMD
static properties with the results of standard PIMD simulations.

\begin{figure}
	\includegraphics{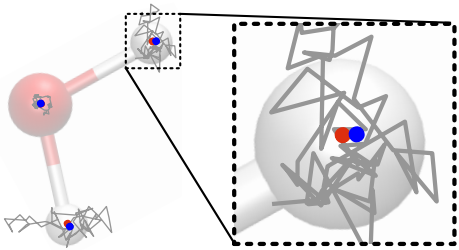}
	\caption{\label{fig:eckart} 
    A typical ring-polymer geometry from a QCMD water simulation at 150~K,
    showing how the Eckart-like conditions  have
    oriented the polymer such that the atom centroids (red) almost coincide
    with the atom quasi-centroids (blue). 
		}
\end{figure}

\subsection{Simulation results%
\label{sec:cond-phase}}
\subsubsection{Computational details}
\label{details}

Condensed-phase water CMD, TRPMD and QCMD simulations were carried out in the
three regimes studied in \sref{trpmd2}, namely the compressed liquid at
$600~\si{K}$, the liquid at $300~\si{K}$, and ice I$_{\rm h}$ at $150~\si{K}$.

\gtnote{Unless noted otherwise}, the CMD and TRPMD calculations used the same parameters as \sref{trpmd2}, with
periodic boundary conditions\cite{frenkel,allen} and simulation boxes of 128~molecules for the
liquid and 96 for ice. At each temperature, a set of independent starting
configurations was generated by propagating eight instances of the system
classically for $100~\si{ps}$ under the action of a global Langevin
thermostat,\cite{schneider-langevin,bussi-langevin} followed by a $100~\si{ps}$ TRPMD run under
PILE-G \gtnote{(with the centroid relaxation time $ \tau_0 = 100~\si{fs}$, and the non-centroid mode friction parameter $ \lambda = 0.5 $)}.\cite{pile} 
The resulting eight equilibrated ring-polymer configurations
were then propagated for a further $100~\si{ps}$\gtnote{, using \mbox{PILE-G} (with the same parameters as above)} to give the TRPMD 
results, and
the PA-CMD algorithm \gtnote{(with $\gamma=4$, $ \tau_0 = 100~\mathrm{fs} $, and $\lambda = 0.01$)} to give the CMD results.

To compute the QCMD spectra for the liquid, we took the eight equilibrated
configurations obtained after the first 100 ps PILE-G (see above), then
propagated for $35~\si{ps}$ using the AQCMD algorithm as described in
\ssecc{sec:aqcmd}{sec:eckart},  with a global Langevin
thermostat\cite{bussi-langevin} (time constant $\tau_0 = 100~\si{fs}$) acting
on the quasi-centroids. The first $10~\si{ps}$ were discarded before
calculating the dipole-derivative time-correlation functions. 

For QCMD ice, we generated 15~independent equilibrated configurations  (in the
same manner as the~8 in the liquid simulations). These were then propagated
with the AQCMD algorithm for $2.5~\si{ps}$, with a local Langevin
thermostat\cite{bussi-langevin} acting on the quasi-centroids. For each of the
15 configurations, the system was then propagated under a global Langevin
thermostat\cite{bussi-langevin} for three intervals of $2.5~\si{ps}$, 
separated by relaxation periods of $300~\si{fs}$, during which local
thermostatting was used. The IR spectrum was then calculated by averaging over
the three $2.5~\si{ps}$ trajectories from the 15 independent simulations.
Converged AQCMD results were obtained using $\gamma = 8$ at 600 K, and $\gamma
= 32$ at 300 K. At 150~K, the calculation was stopped at $\gamma = 128$, which
has definitely converged the positions of the peaks and the $t \to 0$
limit of the dipole-derivative time-correlation function, but may not yet have converged the
intensity of the stretch peak---see \app{app:qtip4pf-conv}. These values of
$\gamma$ made the QCMD calculations approximately $2,\,8,\,\text{and}\ 32$
times more expensive than the corresponding CMD calculations, which used the PA algorithm\cite{hone2,mano-pacmd} and thus 
smaller values of $\gamma$.

\subsubsection{\gtnote{%
Overview of the QCMD spectrum}%
\label{sec:cond-results}}

\begin{figure}
\includegraphics{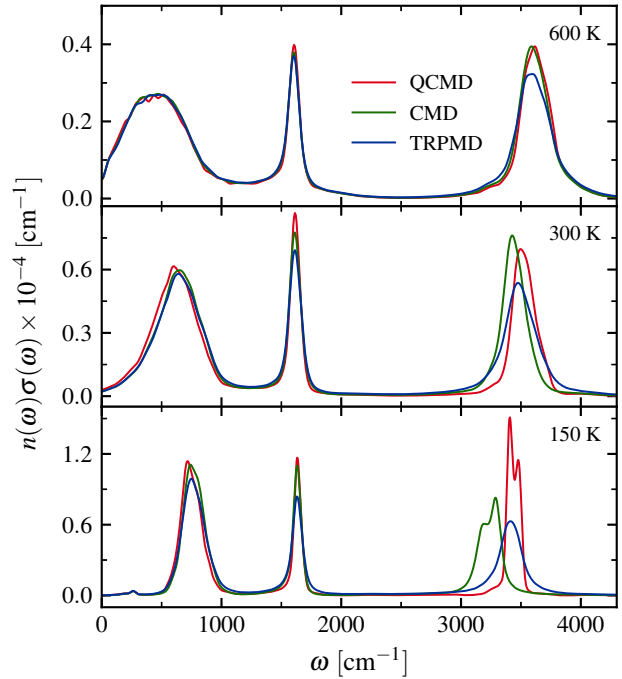}
\caption{\label{fig:qtip4pf-spectra} 
        Simulated infrared absorption spectra for q-TIP4P/F water, at 600~K
        (compressed liquid), 300~K (liquid), and 150~K (ice I$_{\rm h}$),
        obtained by damping the time-correlation function with a Hann window of
        width $\tau=600~\si{fs}$ at 600 and 300~K, and $\tau = 800~\si{fs}$ at
        150~K.  The bond-angle centroids used in the QCMD simulations appear to
        have eliminated the red-shifts and broadening seen in the CMD stretch
        band at 300 and 150~K.
         }
\end{figure}

At 600~K,  the QCMD spectra are in very close agreement with the TRPMD and CMD
spectra (\sfig{fig:qtip4pf-spectra}). Measured from the peak maximum, the QCMD
OH stretch at $3600~\mathrm{cm}^{-1}$ is blue-shifted by about
$15~\mathrm{cm}^{-1}$ relative to CMD and TRPMD. The bend at
$1600~\mathrm{cm}^{-1}$ and the libration peak at $500~\mathrm{cm}^{-1}$ are
almost identical in all three methods.  

At 300~K, noticeable differences emerge between QCMD and the other two methods,
due to the CMD curvature problem, and the broadening of TRPMD, although 
these problems are a lot less severe than in the gas-phase at this
temperature.\cite{marx10,trpmd2}
The CMD stretch peak is red-shifted relative to TRPMD by about
$50~\mathrm{cm}^{-1}$; the QCMD stretch peak is blue-shifted 
by about $20~\mathrm{cm}^{-1}$. The bend peaks are in good agreement for all three
methods. The libration band in QCMD is slightly red-shifted; the most possible
cause for this small shift is the approximation made to the torques on the
monomers in \eqn{eq:qc-torque-approx} (see Sec.~\ref{boo}). 

In $150~\si{K}$ ice, there are major differences between the three sets of
results\gtnote{, mainly in the stretch region}. The curvature problem is strong, artificially broadening and
red-shifting the CMD stretch by more than $100~\mathrm{cm}^{-1}$, and the TRPMD stretch
is severely broadened. In contrast, the QCMD stretch is  sharp, with
resolved symmetric and antisymmetric bands. This
resolution of the stretch bands is an artefact
of the \mbox{q-TIP4P/F} surface and is also observed
in classical simulations. \gtnote{Using a shorter time window of 500~\si{fs} to
coalesce the QCMD stretch peaks (not shown), gives a QCMD stretch} that is blue-shifted by about $10~\mathrm{cm}^{-1}$ with
respect to the TRPMD stretch, and is more than twice as intense. Note that (as
mentioned above) the intensity of the QCMD stretch may not have converged with
respect to $\gamma$.  \gtnote{All three methods are in close agreement in
the libration and bend regions, with a slight red-shift in the QCMD libration band 
which is smaller than the shift at 300 K (see Sec.~\ref{boo}).}

\subsubsection{\gtnote{%
Stretch region of the QCMD infrared spectrum}%
\label{sec:cond-stretch}}

\gtnote{%
   To better assess the performance of QCMD in the stretch region of the spectrum  ($>2000$ cm$^{-1}$), we compare 
   with two other methods: the coloured-noise version of TPRMD developed recently by Rossi et al.\cite{trpmd3} and the local monomer approximation
   (LMon) of Bowman and co-workers.\cite{lmon1,lmon2,trpmd2} Both methods are expected to do well in the stretch region of the spectrum, but to suffer from
   artefacts at lower frequencies.
   }
   
\gtnote{%
In coloured-noise TRPMD, the    white-noise PILE thermostat used in most TRPMD calculations (such as those reported above) is replaced
by a generalised Langevin equation (GLE) thermostat that is designed to minimise 
the dynamical disturbance to the centroids at certain pre-tuned frequencies.\cite{trpmd3}  Using the GLE(C) parametrisation of the thermostat in ref.~\onlinecite{trpmd3} (along with the other simulation parameters in this reference), we propagated eight independent 100~\si{ps} \mbox{TRPMD+GLE(C)} trajectories at 300 and 150 K, using the \mbox{i-PI} package.\cite{ipi} The resulting spectra in the stretch-region are shown
in \sfig{fig:qtip4pf-lmon4}; (the lower-frequency parts of the spectrum, which are corrupted by the thermostat, are shown in \app{app:gle-lmon4}).
}

\begin{figure}
	\includegraphics{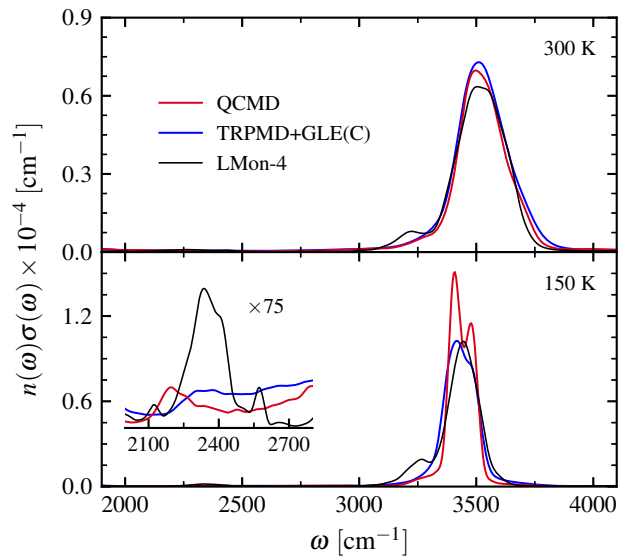}
	\caption{\label{fig:qtip4pf-lmon4} 
        \gtnote{%
		The stretch region of the infrared absorption spectra from the same QCMD simulations
		as \sfig{fig:qtip4pf-spectra}, compared with TRPMD+GLE(C) spectra calculated as in~\sref{trpmd3}, and
	         LMon-4 spectra taken
		from~\sref{trpmd2}.
		The inset magnifies the combination-band part of
		the spectrum. 
        }
	}
\end{figure}
 
\gtnote{%
In the LMon method, the solid or liquid is
equilibrated using standard PIMD techniques, then the Schr\" odinger equation for the nuclear dynamics is solved for each monomer independently, with all degrees of freedom except for
the internal and few intramolecular modes of the monomer held fixed.\cite{lmon1,lmon2,lmon3,lmon4} 
Here, we make use of the results of previous LMon calculations,\cite{trpmd2} in which the PIMD equilibration was done
using the same box size and simulation parameters as in the TRPMD calculations of Sec.~\ref{details},
and the internal modes of the water molecule were coupled to each of its
three librational modes in turn, resulting in a four-mode approximation
(LMon-4). We convolved the raw data obtained from these calculations\cite{mariana} with the same time-windows as used in the QCMD calculations (see \app{app:spectra-calc}), to obtain the spectra shown in \sfig{fig:qtip4pf-lmon4} for the stretch region at 300 and 150 K.
The approximate treatment of the intermolecular modes means that LMon-4 gives a relatively poor description of the spectrum at 600~K (not shown)
and of the libration at all temperatures (see \app{app:gle-lmon4}).}

\gtnote{%
The agreement between the QCMD and TRPMD+GLE(C) stretch peaks is excellent: at 300 K the peaks are almost identical; at 150 K the TRPMD+GLE(C) peaks line up with the QCMD peaks, but are somewhat broader, with the bifurcation just visible; (the broadening is to be expected since the thermostat couples more strongly to the centroid at lower temperatures).  There is probably some cancellation of errors between these two sets of results, as the QCMD method makes a number of static approximations (see \ssec{qcmd2d}), and the GLE(C) thermostat interferes strongly with the dynamics of the librational modes which are coupled to the dynamics of the stretch modes. One piece of evidence for such errors is that the 300 K QCMD peak is slightly less intense than the TRPMD+GLE(C) peak, whereas one would expect the opposite to be true. Nonetheless the agreement between QCMD and TRPMD+GLE(C) in the stretch region is remarkable, suggesting that both methods give excellent approximations to the Matsubara-dynamics spectrum in this region.}

\gtnote{%
       The QCMD and LMon-4 results are also in good agreement in the stretch region (\sfig{fig:qtip4pf-lmon4}). Clearly one cannot use
LMon-4 as a quantum benchmark, since it includes only a few degrees of freedom in the monomer dynamics (which is probably why it does not reproduce the  bifurcation in the stretch peak). Nevertheless, we think the comparison in \sfig{fig:qtip4pf-lmon4} highlights an important weaknesses in the QCMD approach  (which is shared by CMD\cite{paesnew} and TRPMD\cite{trpmd2}). The LMon-4 spectrum gives a libration-bend combination band at roughly $2300~\wn$ (see the \sfig{fig:qtip4pf-lmon4} inset) and a Fermi resonance overlapping with the stretch peak at $3200~\wn$; both these features are much more intense than the corresponding features in the QCMD, CMD and TRPMD spectra. While we cannot say how well the LMon-4 calculations have converged these features, it seems highly likely that QCMD, CMD and TRPMD grossly underestimate them. This is not surprising, as we expect such features to depend  strongly on Matsubara
fluctuations,\cite{mats,mats-rpmd} and also probably on real-time coherence---effects which  QCMD, CMD and TRPMD omit.
}

\subsubsection{\gtnote{%
        Approximations to the quantum Boltzmann distribution}}\label{boo}

\begin{figure}
	\includegraphics{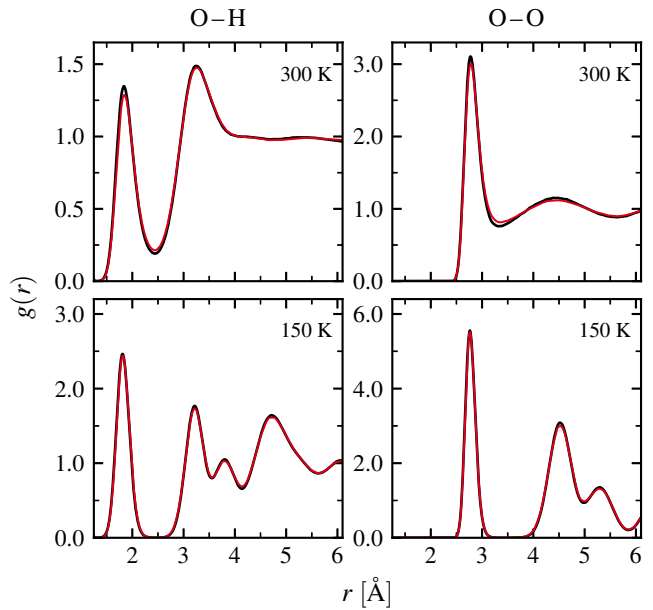}
	\caption{\label{fig:qtip4pf-rdfs}
        \gtnote{%
                Simulated oxygen-hydrogen (\ce{O-H}) and oxygen-oxygen (\ce{O-O}) radial distribution functions (RDFs) for
                \mbox{q-TIP4P/F} water, calculated using QCMD (red) and standard PIMD (black).
                Hydrogen-hydrogen RDFs (not shown) are in similarly close agreement.
                The QCMD results are given for an adiabatic separation of $ \gamma = 64 $ at $300~\si{K}$
                and $\gamma = 128$ at $150~\si{K}$. Only the intermolecular part of the distribution is shown for \ce{O-H}; 
                the intramolecular distributions obtained using the two methods agree to within graphical accuracy.            }
	}
\end{figure}

\gtnote{%
On the basis of the spectra reported above, it seems likely that QCMD gives a similarly good approximation to the quantum Boltzmann distribution for the condensed phase as it does for the two-dimensional model and the gas-phase. However, the condensed-phase QCMD calculations 
make extra approximations to the quantum Boltzmann distribution: \eqn{eq:qc-torque-approx} approximates the torques on the monomers, and the AQCMD algorithm only samples the exact quantum Boltzmann distribution in the limit $\gamma\to\infty$.}

\gtnote{%
To test these static approximations, we compare (in \sfig{fig:qtip4pf-rdfs}) the AQCMD radial distribution functions (RDFs) for the liquid at 300 K and ice at 150 K with the exact RDFs (obtained from PIMD). As expected, the AQCMD RDFs are close to the exact values, although there are small errors at 300 K, indicating that AQCMD gives a slightly less structured liquid  than it should.  Given that the AQCMD RDFs are in almost perfect agreement with the exact values for 150~K ice, and that the intramolecular peaks in the O--H RDF (not shown) agree to graphical accuracy, we think that the (small) errors at 300~K
are mainly the result of approximating the monomer torques by \eqn{eq:qc-torque-approx}. There is perhaps also a very small contribution to the errors  from incomplete convergence in $\gamma$; increasing $\gamma$ from 32 to 64 reduced slightly the errors in the RDFs, suggesting that higher values of
 $\gamma$ might reduce them further.}

\gtnote{These results suggest that QCMD gives an excellent approximation to the quantum Boltzmann distributions in water and ice,
 except for small errors in the long-range correlations at 300~K, which are most likely due to the quasi-centroid torque estimator of
  \eqn{eq:qc-torque-approx}. This estimator is also thought to be responsible for the small red shift in the libration band of the infrared spectrum at 300~K (see \sfig{fig:qtip4pf-spectra}). Future work may thus be able to reduce both of these (small) errors, by developing better
  torque estimators.}

\section{%
\label{sec:conclusions}%
Conclusions}

\gtnote{%
        By making minor approximations to the quantum Boltzmann distribution, we have obtained a method (QCMD) for propagating curvilinear mean-field centroids which is at least efficient enough to treat liquid water and ice on the q-TIP4P/F potential. For a two-dimensional model,
 the quantum Boltzmann approximations are found to be almost exact, and the infrared spectrum is close to that obtained from Matsubara dynamics, showing no sign of the `curvature problem' that afflicts CMD at low temperatures. For liquid water (300 K) and ice (150 K),
the errors made in the quantum Boltzmann approximations are found to be small, suggesting that the QCMD spectra are also very close to the Matsubara-dynamics spectra for these systems. The algorithm we developed to
  implement QCMD is slow (taking 32 times longer than PA-CMD for 150~K ice), but is do-able, and we hope 
  to apply it soon to more realistic water potentials such as MB-pol, to allow comparison with experiment. It may be possible to speed up the algorithm by making it partially adiabatic, or perhaps to develop instead a curvilinear version of TRPMD.}

\gtnote{%
As with other ring-polymer dynamics methods, such as CMD and (T)RPMD, QCMD is a (quasi-)centroid-following method, and thus cannot obtain information about the dynamics of the Matsubara fluctuations around the centroid. In general, QCMD will
thus give poor results for non-linear operators \cite{waterpole} (which depend explicitly on the dynamics of the Matsubara fluctuation modes), and for overtones and combination bands (which couple the centroid strongly to the fluctuation modes)---as an example of the latter, see \sfig{fig:qtip4pf-lmon4}.}

\gtnote{%
How readily can QCMD be generalised to systems other
than water? This question is equivalent to asking how readily collective bead coordinates can be identified, around which the ring-polymers are distributed compactly. Such coordinates are probably easy to find for liquids and solids made up of small, rigid molecules; but further work will be required to determine whether they can be found for systems containing floppy molecular units, such as solvated protons. Also, for reaction barriers below the instanton cross-over temperature, the collective bead coordinates will need to include the unstable mode of the instanton. A `holy grail' would be to develop an 
algorithm that follows, automatically, the collective bead coordinate that minimises the spread of the ring-polymer distribution, thus giving the best mean-field approximation to Matsubara dynamics.}

\begin{acknowledgments}
        \gtnote{%
We thank Mariana Rossi for sending us the LMon-4 data used to obtain \sfig{fig:qtip4pf-lmon4}.}
G.T.\ acknowledges a University of Cambridge Vice-Chancellor's award and
support from St.~Catharine's College, Cambridge. 
M.J.W.\ and S.C.A.\ acknowledge funding from the UK Science and Engineering Research
Council.
\end{acknowledgments}

\begin{appendix}

        \section{\gtnote{%
                Details of the calculations of the infrared absorption spectra}%
\label{app:spectra-calc}}

The champagne-bottle potential used in \ssec{sec:2d-morse} is
\begin{align}
\label{eq:morse}
        V(r)=D_0 \left[
                1 - e^{-\alpha (r - r_{\mathrm{eq}})}
                 \right]^{2},
\end{align}
where $r = \big(q_x^2 + q_y^2\big)^{1/2},\, r_\mathrm{eq} = 1.8324$,
$D_0 = 0.18748$, and $\alpha = 1.1605$ a.u. 
The mass of the particle is taken to be $m=1741.1$ a.u.

        \gtnote{%
For the linear dipole-moment calculations (i.e.~the 2D model and the condensed-phase water calculations), the infrared spectrum was obtained  from the Kubo-transformed time-autocorrelation function (TAF) $ C(t) $ of the cell dipole-moment derivative~$\dot{\bm{\mu}}$\cite{tfmiller1,mano-pacmd}
\begin{gather}
    {
    \tilde{I}_{\dot{\bm{\mu}}}(\omega) = \int_{-\infty}^\infty\!dt\,
    e^{-i\omega t} C_{\dot{\bm{\mu}}\cdot\dot{\bm{\mu}}}(t) f(t) 
    } \no \\
    \label{eq:golden-rule2}
    {
      n(\omega)\sigma(\omega) = \dfrac{\beta}{6 c V \!\epsilon_0} \tilde{I}_{\dot{\bm{\mu}}}(\omega),
    }
\end{gather}
where $ V $ is the cell volume, $ n(\omega) $ is the refractive index, $ \sigma(\omega) $ is the absorption cross-section, and $f(t)$ is a window function that dampens the tail of the TAF,
reducing the ringing artefacts.\cite{nr77} For the gas-phase water calculations (which employed the Partridge-Schwenke dipole-moment surface), we used the dipole TAF
\begin{gather}
        {
        \tilde{I}_{\bm{\mu}}(\omega) = \int_{-\infty}^\infty\!dt\,
         e^{-i\omega t} C_{\bm{\mu}\cdot\bm{\mu}}(t) f(t)
        }\no \\
        \label{eq:golden-rule}
        {
        n(\omega)\sigma(\omega) = \dfrac{\beta}{6 c V \!\epsilon_0} \, \omega^2 \tilde{I}_{\bm{\mu}}(\omega).
        }
\end{gather}
In two-dimensional and gas-phase calculations, where the cell volume $ V $ is not defined,
   we take $ \tilde{I}_{\dot{\bm{\mu}}}(\omega) $ or $\omega^2 \tilde{I}_{{\bm{\mu}}}(\omega) $ to be ``the spectrum'' instead of $ n(\omega) \sigma(\omega) $.}

When applied to the exact quantum TAF, \gtnote{the damping envelope $ f(t) $ serves the additional purpose of removing the recurrences that are caused by the real-time quantum coherence.  In the two-dimensional model, the recurrences} are well
separated from the initial part of the TAF, allowing a sigmoid window with a
sharp cut-off to be used, 
\begin{align}\label{twin}
        f(t)=\dfrac{1}{
                1 + e^{(|t| - t_{1/2})/\tau}
                },
\end{align}
with parameters  $t_{1/2} = 400~\si{fs},\, \text{and}\ \tau = 25~\si{fs}$. 
In simulations of gaseous and condensed-phase water
(\ssecc{sec:ps-h2o}{sec:cond-phase}) the Hann window\cite{nr77}
        \gtnote{%
\begin{align}
f(t) = \begin{cases}
          \cos^2\!\left( \frac{\pi t}{2 \tau} \right) & |t| \leq \tau \\
          0                                  & |t| > \tau
       \end{cases} \label{eq:hann}
\end{align}
is found to be more suitable. 
The cut-off time $ \tau $ is set to 600~\si{fs} for liquid water at 600 and 300~\si{K}, and to 800~\si{fs} for ice at 150~\si{K}.}

\gtnote{%
To obtain smooth LMon-4 spectra, for which the raw data\cite{mariana} are a list of discrete transition frequencies $ \{\omega_i\} $, we calculate
	\begin{align}
	&  n(\omega) \sigma(\omega) \propto \sum_{i} \mu_i \tilde{f}(\omega - \omega_i) \nonumber \\
	&  \tilde{f}(\omega) = \dfrac{\sin(\omega \tau)}{\omega[1 - (\omega \tau / \pi)^2]},
	\end{align}
where $ \mu_i $ are the corresponding transition dipole moments and $ \tilde{f} $ is the Fourier transform of the Hann window. The resulting spectrum is then scaled so that its area integrated
between  2600 and 4500~\wn agrees with QCMD.}

\section{Treating the quasi-centroid momentum as purely cartesian%
\label{app:qc-approx}}

To obtain \eqn{eq:2df}, we assume that the quasi-centroid momentum $\overline {\bf P}$ contributes a cartesian term $e^{-\beta\overline{\bf P}^2/ 2m}$ to the
mean-field-averaged partition function $Z_p(P_R,L,R,\Theta)$, where $P_R$ and $L$ are the momenta conjugate to $R$ and $\Theta$. To investigate this approximation, it helps to specify $\Theta$. Let us use
\begin{align}
\Theta=\dfrac{1}{N}\sum_i\theta_i
\end{align}
with $\theta_i=\arctan{y_i/x_i}$. We then obtain (without approximation)
\begin{align}
        Z_p(P_R,L,R,\Theta) & = \int\!\int\! \dfrac{J({\bf r}')}{\rho'} d{\bf r}'d{\bm \theta}'\no\\
        {} & \times e^{-\beta \left[T(P_R,L,\rho')+W({\bf r}',{\bm \theta}')\right]}\no\\
        {} & \times\delta\!\left( R'- R\right)\delta\!\left( \Theta'- \Theta\right)%
\label{eq:2d-part-exact}
\end{align}
with
\begin{align}\label{bjac}
        T(P_R,L,\rho) & = {P_R^2\over 2m}+{L^2\over 2m\rho^2}\no\\
        \rho & = \sqrt{\dfrac{1}{N} \sum_i r_i^2}\no\\
        J({\bf r}) & = \prod_{i=1}^Nr_i
\end{align}
To make $T(P_R,L,\rho)$ factorise out of $Z_p(P_R,L,R,\Theta)$ we need to approximate the moment of inertia $2m\rho^2$ by $2mR^2$, giving
\begin{align}
T(P_R,L,\rho)\simeq{P_R^2\over 2m}+{L^2\over 2m R^2}={\overline{\bf P}^2\over 2m}
\end{align}
and to make the corresponding change $1/{\rho}'\to 1/R'$ in the integrand of \eqn{eq:2d-part-exact}. This gives
\begin{align}
        Z_p(P_R,L,R,\Theta) & \simeq e^{-\beta\overline{\bf P}^2/ 2m} \int\!\int\! {J({\bf r}')\over R'}d{\bf r}'d{\bm \theta}'\no\\
        {} & \times e^{-\beta W({\bf r}',{\bm \theta}')}\no\\
        {} & \times\delta\!\left( R'- R\right)\delta\!\left( \Theta'- \Theta\right)\no\\
           & = e^{-\beta\overline{\bf P}^2/ 2m}{{\overline Z}_0(R)\over 2\pi R}
\end{align}
Since 
\begin{align}
\rho^2=R^2\left[1+{1\over N}\sum_i{{\Delta r_i}^2\over R^2} \right]
\end{align}
where $\Delta r_i=r_i-R$, it follows that treating $\overline{\bf P}$ as purely cartesian is a good approximation if the 
ring-polymer distribution is sufficiently compact that we can neglect terms quadratic in ${\Delta r_i}$ in the moment of inertia in $T(P_R,L,\rho)$.


\section{Convergence of AQCMD spectra for liquid water and ice%
\label{app:qtip4pf-conv}}

The convergence of the AQCMD infrared absorption spectra with respect to the adiabatic
separation $\gamma$ at 600 and 300~K was tested by simulating liquid water 
using a box of 32 molecules, subject to periodic boundary conditions.\cite{frenkel,allen}
The simulated spectra converged at $\gamma = 8\ \text{and}\ 32$ respectively, as
shown in \fig{fig:qtip4pf-conv}.

\begin{figure}[h]
	\includegraphics{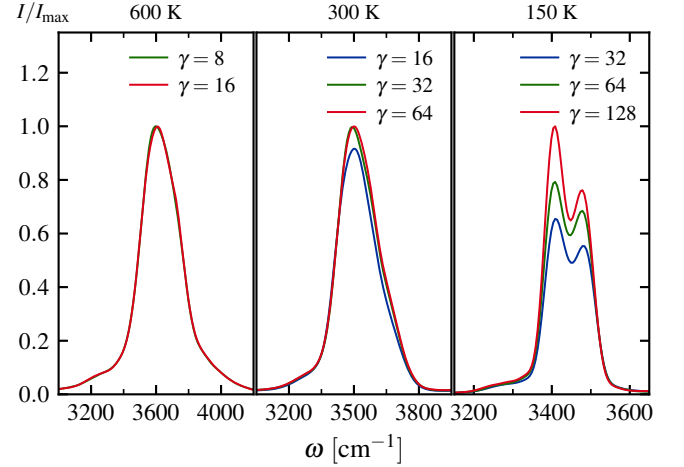}
	\caption{\label{fig:qtip4pf-conv} 
		Simulated infrared absorption spectra in the OH-stretch region of
		q-TIP4P/F water at $600$, $300$, and 150~K, calculated with the 
		AQCMD algorithm of \ssec{sec:aqcmd} at
		different levels of adiabatic separation~$\gamma$. In all cases the
		position of the absorption maximum remains largely unchanged with
		increasing $\gamma$, and only the intensity is noticeably affected. The
		spectrum at 300~K demonstrates that convergence of the intensity may be
		quite abrupt, and similar behaviour is anticipated at $150~\si{K}$.
	}
\end{figure}

The convergence with respect to $\gamma$ for ice at 150~K was tested using a 
larger simulation box of 96~molecules (in order to give zero net dipole moment\cite{hayward}).
The highest value tested was $\gamma = 128$, which was sufficient to converge the positions of all the bands in the spectrum,
but not the relative intensity of the bend and stretch bands (see \sfig{fig:qtip4pf-conv}). A value of
$\gamma = 64$ was sufficient to converge the dipole-derivative TAF at  $t = 0$. This,
together with the abrupt nature of convergence at $300~\si{K}$,
suggests that the $\gamma=128$  spectrum is close to convergence at 150~K.

\section{\gtnote{%
        LMon-4 and TRPMD+GLE spectra for water}%
	\label{app:gle-lmon4}}
\begin{figure}[h]
	\includegraphics{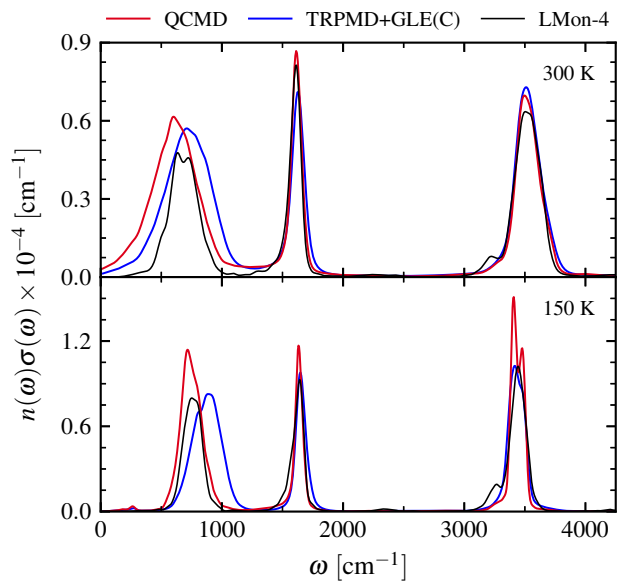}
	\caption{\label{fig:lmon4-full} 
        \gtnote{%
                Infrared absorption spectra from the same simulations as 
            \sfig{fig:qtip4pf-lmon4}, showing also the libration and bend regions.
            }
	}
\end{figure}

\gtnote{%
For completeness we compare in \sfig{fig:lmon4-full} the full IR spectra calculated using the QCMD, TRPMD+GLE(C) and LMon-4 methods
 (the stretch regions of the spectra are also plotted in \sfig{fig:qtip4pf-lmon4} and discussed in \ssec{sec:cond-stretch}). As mentioned in  \ssec{sec:cond-stretch}, LMon-4 and TRPMD+GLE(C) are known to provide poor descriptions of the libration band.}

\end{appendix}
\nocite{*}

\begin{thebibliography}{10}

\bibitem{skinner2009} J.L.~Skinner, B.M.~Auer, and Y.-S.~Lin, 
        \textit{Advances in Chemical Physics} (Wiley-Blackwell, 2009), pp.~59--103.

\bibitem{skinner2013} S.M.~Gruenbaum, C.J.~Tainter, L.~Shi, Y.~Ni, and J.L.~Skinner, 
        \jctc {\bf 9}, 3109 (2013).

\bibitem{lmon1} H.~Liu, Y.~Wang, and J.M.~Bowman, J.\ Am.\ Chem.\ Soc.\ {\bf 136}, 5888 (2014).

\bibitem{lmon2} H.~Liu, Y.~Wang, and J.M.~Bowman, \jpcb {\bf 118}, 14124 (2014).

\bibitem{billjpc} W.H.~Miller, \jpca {\bf 105}, 2942 (2001).

\bibitem{liubill}  J.~Liu and W.H.~Miller, \jcp {\bf 131}, 074113 (2009).

\bibitem{liurev} J.~Liu, Int.\ J.\ Quantum Chem.\ {\bf 115}, 657 (2015).

\bibitem{shig} Q.~Shi and E.~Geva, \jpca {\bf 107}, 9059 (2003).

\bibitem{jens1} J.A.~Poulsen, G.~Nyman and P.J.~Rossky, \jcp {\bf 119}, 12179 (2003).

\bibitem{xantheas} J.~Liu, W.H.~Miller, G.S.~Fanourgakis, S.S.~Xantheas, S.~Imoto, and S.~Saito, \jcp {\bf 135}, 244503 (2011).

\bibitem{liuwat} X.~Liu and J.~Liu, Mol.\ Phys.\ {\bf 116}, 755 (2018).

\bibitem{bonella} E.~Mangaud, S.~Huppert, T.~Pl{\' e}, P.~Depondt, S.~Bonella, and F.~Finocchi, \jctc \textbf{15}, 2863 (2019).

\bibitem{vollume} M.~Basire, D.~Borgis, and R.~Vuilleumier, \pccp {\bf 15}, 12591 (2013).

\bibitem{van1} J.~Van{\' i}{\v c}ek, W.H.~Miller, J.F.~Castillo, and F.J.~Aoiz, \jcp {\bf 123}, 054108 (2005).

\bibitem{van2} J.~Van{\' i}{\v c}ek and W.H.~Miller, \jcp {\bf 127}, 114309 (2007).

\bibitem{jerin1} J.O.~Richardson, \jcp {\bf 144}, 114106 (2016).

\bibitem{jerin2} J.O.~Richardson, \jcp {\bf 148}, 200901 (2018).

\bibitem{caov2} J.~Cao and G.A.~Voth, \jcp {\bf 100}, 5106 (1994).

\bibitem{craig1} I.R.~Craig and D.E.~Manolopoulos, \jcp {\bf 121}, 3368 (2004).

\bibitem{craig2} I.R.~Craig and D.E.~Manolopoulos, \jcp {\bf 122}, 084106 (2005).

\bibitem{tfmiller1} \gtnote{T.F.~Miller and D.E.~Manolopoulos, \jcp {\bf 123}, 154504 (2005).}

\bibitem{annurev} S.~Habershon, D.E.~Manolopoulos, T.E.~Markland, and T.F.~Miller~III,
 Annu.\ Rev.\ Phys.\ Chem.\ {\bf 64}, 387 (2013).

\bibitem{trpmd1} M.~Rossi, M.~Ceriotti, and D.E.~Manolopoulos, \jcp {\bf 140}, 234116 (2014).

\bibitem{jorsca} J.O.~Richardson and S.C.~Althorpe, \jcp {\bf 131}, 214106 (2009).

\bibitem{tim} T.J.H.~Hele and S.C.~Althorpe, \jcp {\bf 138}, 084108 (2013).

\bibitem{tom1} N.~Boekelheide, R.~Salom\' on-Ferrer, and T.F.~Miller III, \pnas {\bf 108}, 16159 (2011).

\bibitem{tom2} J.S.~Kretchmer and T.F. Miller III, \jcp {\bf 138}, 134109 (2013).

\bibitem{paes1} G.R.~Medders and F.~Paesani, \jctc {\bf 11}, 1145 (2015).

\bibitem{paes2} S.K.~Reddy, D.R.~Moberg, S.C.~Straight, and F.~Paesani, \jcp {\bf 147}, 244504 (2017).

\bibitem{chanwol} D.~Chandler and P.G.~Wolynes, \jcp {\bf 74}, 4078 (1981).

\bibitem{pariram} M.~Parrinello and A.~Rahman, \jcp {\bf 80}, 860 (1984).

\bibitem{ceperley} D.M.~Ceperley, Rev.\ Mod.\ Phys.\ {\bf 67}, 279 (1995).


\bibitem{marx10} S.D.~Ivanov, A.~Witt, M.~Shiga, and D.~Marx, \jcp {\bf 132}, 031101 (2010).

\bibitem{marx09} A.~Witt, S.D.~Ivanov, M.~Shiga, H.~Forbert, and D.~Marx, \jcp {\bf 130}, 194510 (2009).

\bibitem{trpmd2} M.~Rossi, H.~Liu, F.~Paesani, J.~Bowman, and M.~Ceriotti, \jcp {\bf 141}, 181101 (2014).

\bibitem{trpmd3} M.~Rossi, V.~Kapil, and M.~Ceriotti, \jcp {\bf 148}, 102301 (2018).

\bibitem{mats} T.J.H.~Hele, M.J.~Willatt, A.~Muolo, and S.C.~Althorpe, \jcp \textbf{142}, 134103 (2015).

\bibitem{mf-mats} G.~Trenins and S.C.~Althorpe, \jcp \textbf{149}, 014102 (2018).

\bibitem{jens2} K.K.G.~Smith, J.A.~Poulsen, G.~Nyman, and P.J.~Rossky, \jcp {\bf 142}, 244112 (2015).

\bibitem{planets} M.J.~Willatt, M.~Ceriotti, and S.C.~Althorpe, \jcp {\bf 148}, 102336 (2018).

\bibitem{mats-rpmd} T.J.H.~Hele, M.J.~Willatt, A.~Muolo, and S.C.~Althorpe, \jcp \textbf{142}, 191101 (2015).

\bibitem{miller-ham} W.H.~Miller, N.C.~Handy, and J.E.~Adams, \jcp {\bf 72}, 99 (1980).

\bibitem{tennyson-ham} B.T.~Sutcliffe and J.~Tennyson, Mol.\ Phys. {\bf 58}, 1053 (1986).

\bibitem{truhlar-ham} C.F.~Jackels, Z.~Gu, and D.G.~Truhlar, \jcp {\bf 102}, 3188 (1995).

\bibitem{marx-curv} D.~Marx and M.H.~M\"{u}ser, J.\ Phys.: Condens.\ Matter {\bf 11}, R117 (1999). 

\bibitem{kleinert} H.~Kleinert, \textit{Path Integrals in Quantum Mechanics, Statistics, Polymer Physics, and Financial Markets} (World Scientific, 2009),
        pp.~697--751. 



\bibitem{caov4} J.~Cao and G.A.~Voth, \jcp {\bf 101}, 6168 (1994).

\bibitem{hone1} T.D.~Hone and G.A.~Voth, \jcp {\bf 121}, 6412 (2004).

\bibitem{hone2} T.D.~Hone, P.J.~Rossky, and G.A.~Voth, \jcp {\bf 124}, 154103 (2006).

\bibitem{ryckaerts} J.-P.~Ryckaert, G.~Ciccotti, and H.J.C.~Berendsen, J.\ Comput.\ Phys.\ {\bf 23}, 327 (1977).

\bibitem{andersen} H.C.~Andersen, J.\ Comput.\ Phys.\ {\bf 52}, 24 (1983).

\bibitem{tuckerman} M.~Tuckerman, \textit{Statistical Mechanics: Theory and Molecular Simulation} (OUP, Oxford, 2010).

\bibitem{eckart} C.~Eckart, Phys.\ Rev.\ \textbf{47}, 552 (1935).

\bibitem{bright-wilson} E.B.~Wilson, J.C.~Decius, and P.C.~Cross, \textit{Molecular Vibrations: The Theory of Infrared and Raman Vibrational Spectra} 
        (Dover Publications, 1980), pp.~11--13.

\bibitem{reduced} A prefactor of $N\left({m N}/{2\pi\beta\hbar^2} \right)^{N-1}$ has been cancelled out.

\bibitem{instantons} No thermally accessible instanton cross-over radius has been found in the $R$-constrained ring-polymer distributions.

\bibitem{Jacob} The spring force is equal to minus the gradient of \mbox{$S({\bf r})-1/\beta \ln [J({\bf r})/R]$}, where $J({\bf r})$ is the Jacobian defined in \eqn{bjac}.


\bibitem{pspes} H.~Partridge and D.W.~Schwenke, \jcp \textbf{106}, 4618 (1997).

\bibitem{psdms} D.W.~Schwenke and H.~Partridge, \jcp \textbf{113}, 6592 (2000).

\bibitem{nr77} W.H.~Press, S.A.~Teukolsky, B.P.~Flannery, and W.T.~Vetterling, \textit{
Numerical Recipes in FORTRAN 77} (Cambridge University Press, 1992).

\bibitem{dvr3d} J.~Tennyson, M.A.~Kostin, P.~Barletta, G.J.~Harris, O.L.~Polyansky, J.~Ramanlal, and N.F.~Zobov, \cpc \textbf{163}, 85 (2004).

\bibitem{paes2008} F.~Paesani and G.A.~Voth, \jcp {\bf 129}, 194113 (2008).

\bibitem{coalson} R.D.~Coalson, \jcp \textbf{85}, 926 (1986).

\bibitem{leimkuhler} \gtnote{B.~Leimkuhler and C.~Matthews, Proc.\ R.\ Soc.\ A \textbf{472}, 20160138 (2016).}

\bibitem{pile} M.~Ceriotti, M.~Parrinello, T.E.~Markland, and
D.E.~Manolopoulos, \jcp \textbf{133}, 124104 (2010).

\bibitem{mano-pacmd} S.~Habershon, G.S.~Fanourgakis, and D.E.~Manolopoulos, \jcp \textbf{129}, 074501 (2008).

\bibitem{trpmd2-sup} See the supplementary material for ref.~\onlinecite{trpmd2}.


\bibitem{qtip4pf} S.~Habershon, T.E.~Markland, and D.E.~Manolopoulos, \jcp {\bf 131}, 024501 (2009).

\bibitem{jorg} F.~J{\o}rgensen, Int.\ J.\ Quantum Chem.\ \textbf{14}, 55 (1978).

\bibitem{kudin} K.N.~Kudin and A.Y.~Dymarsky, \jcp \textbf{122}, 224105 (2005).

\bibitem{minsp}In the absence of an external torque, the Eckart-like condition minimises the spread of the ring-polymer distribution around the orientational angles.

\bibitem{invert} To evaluate $\overline{\bm \tau}$ exactly we would need to express it in terms of cartesian bead coordinates by inverting the nine conditions imposed by \eqnn{eq:h2o-theta}{eckboth}.


\bibitem{frenkel} D.~Frenkel and B.~Smit, \textit{Understanding Molecular
        Simulation: From Algorithms to Applications} (Elsevier Science, 2001).


\bibitem{allen} M.P.~Allen and D.J.~Tildesley, 
        \textit{Computer Simulation of Liquids} 
        (Oxford Science Publications, 1989).

\bibitem{schneider-langevin} T.~Schneider and E.~Stoll, Phys.\ Rev.\ B
        \textbf{17}, 1302 (1978).

\bibitem{bussi-langevin} G.~Bussi and M.~Parrinello, \cpc \textbf{179}, 26
        (2008).
        
\bibitem{ipi} V.~Kapil et al., \cpc \textbf{236}, 214 (2019). 
        
\bibitem{lmon3} Y.~Wang and J.M.~Bowman, \jcp \textbf{134}, 154510 (2011).

\bibitem{lmon4} Y.~Wang and J.M.~Bowman, \jcp \textbf{136}, 144113 (2012).

\bibitem{mariana} M.~Rossi, personal communication (2019).

\bibitem{paesnew} K.M.~Hunter, F.A.~Shakib, and F.~Paesani, \jpcb \textbf{122}, 10754 (2018).

\bibitem{waterpole} \gtnote{%
                Polarisable water dipole-moment surfaces are highly non-linear functions of position, but despite this, classical, CMD and TRPMD simulations of the peak intensities show good overall agreement with experiment. This implies that the main function of the non-linearity is to provide an inhomogenous distribution of local dipole moments, each of which fluctuates almost linearly (although a small Matsubara contribution cannot be ruled out). We thus expect similar behaviour (although with better line shapes and positions) for any future QCMD calculations using these surfaces.}
        

\bibitem{hayward} J.A.~Hayward and J.R.~Reimers, \jcp \textbf{106}, 1518 (1997).

 \end{thebibliography}

\end{document}